\documentclass[amsfonts,amssymb]{article}
\pdfoutput=1
\usepackage{graphicx}

\usepackage{amssymb}
\usepackage[titletoc]{appendix}
\usepackage{color}
\usepackage{subcaption}
\renewcommand{\theequation}{\arabic{equation}}
\newcommand{\be}{\begin{equation}}
\newcommand{\ee}{\end{equation}}
\newcommand{\bea}{\begin{array}}
\newcommand{\ea}{\end{array}}
\newcommand{\beqa}{\begin{eqnarray}}
\newcommand{\eeqa}{\end{eqnarray}}
\newcommand{\bean}{\begin{eqnarray*}}
\newcommand{\eean}{\end{eqnarray*}}

\def\up#1{\leavevmode \raise.16ex\hbox{#1}}

\newcommand{\gapproxeq}{\lower
 .7ex\hbox{$\;\stackrel{\textstyle >}{\sim}\;$}}
\newcommand{\lapproxeq}{\lower .7ex\hbox{$\;\stackrel
{\textstyle <}{\sim}\;$}}

\renewcommand{\theequation}{\thesection.\arabic{equation}}

\newcounter{appendice}
\newcommand{\appendice}
{
\setcounter{equation}{0}
\renewcommand{\theequation}{\Alph{appendice}.\arabic{equation}}
\addtocounter{appendice}{1}

{\Large{\bf  Appendix \Alph{appendice}}}
}

\def\thebibliography#1{{\bf REFERENCES\markboth
 {REFERENCES}{REFERENCES}}\list
 {[\arabic{enumi}]}{\settowidth\labelwidth{[#1]}\leftmargin\labelwidth
 \advance\leftmargin\labelsep
 \usecounter{enumi}}
 \def\newblock{\hskip .11em plus .33em minus -.07em}
 \sloppy
 \sfcode`\.=1000\relax}

\def\BI{{\rm 1\!l}}

\begin{document}

\centerline{ \LARGE Signature change in matrix model solutions }

\vskip 2cm
\centerline{ A. Stern\footnote{astern@ua.edu}  and Chuang Xu\footnote{cxu24@crimson.ua.edu} }
\vskip 1cm
\begin{center}
{ Department of Physics, University of Alabama,\\ Tuscaloosa,
Alabama 35487, USA\\}
\end{center}
\vskip 2cm
\vspace*{5mm}
\normalsize
\centerline{\bf ABSTRACT}

Various classical solutions to  lower dimensional IKKT-like Lorentzian  matrix models are examined in their commutative limit.  Poisson manifolds emerge in this limit, and  their  associated induced  and  effective metrics are computed.  Signature change is found to be a common feature of these manifolds when  quadratic  and cubic terms are included in the bosonic action.  In fact, a single manifold may exhibit multiple signature changes.
Regions with Lorentzian signature may serve as toy models for cosmological space-times, complete with cosmological singularities, occurring at the signature change.  The singularities are resolved away from  the commutative limit.  Toy models of open and closed  cosmological space-times are given in two and four dimensions.  The four dimensional cosmologies  are constructed from non-commutative complex projective spaces, and they are found to display a rapid expansion near the initial singularity.  

\bigskip
\bigskip

\newpage
%\tableofcontents

\section{Introduction}
Signature change is believed to be a feature of quantum gravity.\cite{Atkatz:1981tk}-\cite{Bojowald:2016vlj} 
It has been discussed in the  context of  string theory,\cite{Mars:2000gu} loop quantum gravity,\cite{Mielczarek:2012pf},\cite{Bojowald:2016vlj} and  causal dynamical triangulation\cite{Ambjorn:2015qja}.  Recently, signature change has also been shown to result from for certain  solutions to matrix equations.\cite{Chaney:2015mfa},\cite{Chaney:2016npa},\cite{Steinacker:2017vqw}   These are the classical equations of motion that follow from  Ishibashi, Kawai, Kitazawa and  Tsuchiya (IKKT)-type models,\cite{Ishibashi:1996xs}   with a Lorentzian background  target metric.  The  signature change occurs in the induced  metrics of the continuous manifolds that emerge upon taking the commutative (or equivalently, continuum or semi-classical)  limits of the  matrix model solutions.  Actually, as argued by Steinacker, the  relevant metric  for these emergent manifolds is not, in general, the induced metric, but rather, it is the metric  that appears upon the coupling to matter.\cite{Steinacker:2010rh}  The latter is the so-called effective metric of the emergent manifold, and it is determined from the symplectic structure that appears in the commutative limit, as well as the induced metric. 
Signature changes also occur for the effective metric of these manifolds, and in fact, they precisely coincide with the signature changes in the induced metric. 
The signature changes in the induced or effective metric correspond to singularities in the curvature tensor constructed from these metrics.
The singularities are resolved away from the commutative limit, where the description of the solution is in terms of representations of  some matrix algebras.

As well as being of  intrinsic interest, signature changing matrix model  solutions  could prove useful  for cosmology.
It has been shown that toy cosmological models can be constructed for regions of the manifolds where the metric has  Lorentzian signature.  These regions can represent both open and closed cosmologies,  complete with cosmological singularities which occur at the signature changes.   As stated above such singularities  are  resolved away from the commutative limit.  Furthermore, in \cite{Steinacker:2017vqw}, a rapid expansion, although not exponential, was found to occur immediately after the big bang singularity.

  The previous examples of  matrix models where signature change was observed include the fuzzy sphere embedded in a three-dimensional Lorentzian background,\cite{Chaney:2015mfa}  fuzzy $CP^2$  in an eight-dimensional Lorentzian background,\cite{Chaney:2016npa} and non-commutative $H^4$ in ten-dimensional Lorentz space-time.\cite{Steinacker:2017vqw}   For the purpose of examining  signature changes, it is sufficient to restrict to the bosonic sector of the matrix models.  In this article we present multiple additional examples of  solutions to  bosonic matrix models that exhibit signature change.  We argue that signature change  is actually a common feature of  solutions to IKKT-type matrix models with indefinite background metric, in particular,  when mass  terms  are included in the matrix model action.  (Mass terms have been shown to result from an IR regularization.\cite{Kim:2011cr})  In fact, a single solution can exhibit multiple signature changes.  As an aside, it is known that there are  zero mean curvature surfaces in three-dimensional Minkowski space that change from being space-like to being time-like.\cite{cgu}-\cite{Fujimori2015}\footnote{ We thank J. Hoppe for bringing this to our attention.}  The signature changing surfaces that emerge from solutions to the Lorentzian  matrix models studied here do not have zero mean curvature.

As a warm up, we review two-dimensional  solutions to the three-dimensional Lorentz matrix model, consisting of a quartic (Yang-Mills)  and a cubic term, and without a  quadratic (mass) term. Known     solutions are non-commutative  $(A)dS^2$\cite{Ho:2000fy}-\cite{Pinzul:2017wch}   and  the  non-commutative cylinder.\cite{Chaichian:2000ia},\cite{Balachandran:2004yh},\cite{Stern:2014uea}    They  lead to a  fixed signature upon taking the commutative limit. New solutions appear when the quadratic term is included in the action.  These new solutions exhibit signature change. One such solution, found previously, is the Lorentzian fuzzy sphere.\cite{Chaney:2015mfa} Others can be constructed by deforming  the non-commutative $AdS^2$   solution.  After taking  the commutative limit of these matrix solutions, one finds regions of the emergent manifolds where the metric has  Lorentzian signature. In the case of the Lorentzian fuzzy sphere, the Lorentzian region  crudely describes   a two-dimensional closed cosmology, complete with an initial and final singularity.  In the case of the deformation of non-commutative (Euclidean)  $(A)dS^2$,  the Lorentzian region   describes   a two-dimensional open cosmology.

   Natural extensions of these solutions to higher dimensions are the non-commutative complex projective spaces.\cite{Nair:1998bp}-\cite{Balachandran:2005ew}
Since we  wish to recover  noncompact   manifolds, as well as compact manifolds in the commutative limit, we should consider the indefinite versions  of these non-commutative spaces,\cite{Hasebe:2012mz} as well as those constructed from compact groups.
For four dimensional solutions, there are then three such candidates: non-commutative $CP^2$, $CP^{1,1}$ and $CP^{0,2}$.  The latter two solve an eight-dimensional (massless) matrix model with indefinite background metric, specifically, the $su(2,1)$ Cartan-Killing metric. These solutions  give a fixed signature after taking the commutative limit.  So as with the previous examples,  the massless matrix model yields no signature change.
Once again,  new solutions appear when a mass term is included, and they exhibit signature change, possibly multiple signature changes.  These  solutions include deformations of non-commutative $CP^2$, $CP^{1,1}$ and $CP^{0,2}$.\footnote{As stated above, we assume  to the background metric to be the $su(2,1)$ Cartan-Killing metric. 
 Non-commutative $CP^2$, and its deformations, were shown to solve an eight-dimensional Lorentzian matrix model with a different background metric in \cite{Chaney:2016npa}.}  A   deformed  non-commutative   $CP^{0,2}$ solution can  undergo two signature changes, while a deformed non-commutative $CP^2$ solution can have up to three signature changes.
Upon taking the commutative limit, the deformed  non-commutative  $CP^{1,1}$ and $CP^{0,2}$ solutions  have regions with Lorentzian signature that describe expanding open space-time cosmologies, complete with a big bang singularity occurring at the signature change. The commutative limit of the deformed non-commutative $CP^2$ solution has a  region with Lorentzian signature that describes a closed space-time cosmology, complete with  initial/final singularities.  Like the non-commutative $H^4$ solution found in \cite{Steinacker:2017vqw}, these solutions display extremely rapid expansion near the cosmological singularities.  Also as in \cite{Steinacker:2017vqw},  the space-times emerging from the deformed  non-commutative  $CP^{1,1}$ and $CP^{0,2}$  solutions expand linearly  at late times.  It suggests that these are universal properties of   $4d$ signature changing solutions to IKKT-type matrix models.

The outline for this article is the following:  In section two we review the non-commutative  $(A)dS^2$  and  cylinder solutions to the (massless) three-dimensional Lorentz matrix model.  
We include the mass term  the matrix model action in section three, and examine the resulting signature changing matrix model solutions.  The non-commutative  $CP^{1,1}$ and $CP^{0,2}$ solutions to an eight-dimensional (massless) matrix model (in the semi-classical limit) are  examined   in  section four. The mass term is added to the  action in section five, and there we study the resulting  deformed  non-commutative  $CP^{1,1}$, $CP^{0,2}$ and  $CP^2$ solutions. 
In appendix A we list some properties of $su(2,1)$ in the defining representation.  In appendix B we review a derivation of the effective metric, and compute  it for the examples of
   $CP^{1,1}$ and  $CP^{0,2}$.

\section{   Three-dimensional Lorentzian matrix model }
\setcounter{equation}{0}

We begin by  considering the bosonic sector of  the three-dimensional  Lorentzian matrix model with an action   consisting
of a quartic  (Yang-Mills) term and a cubic term:
\be S(X)=\frac 1{{\tt g}^2}{\rm Tr}\Bigl(-\frac 14 [X_\mu, X_\nu] [X^\mu,X^\nu] +\frac i3  a\, \epsilon_{\mu\nu\lambda}X^\mu[ X^\nu, X^\lambda]\Bigr)\;\label{mmactn3d}\ee
Here $X^\mu,\;\mu=0,1,2$,  are infinite-dimensional hermitean matrices and $a$ and ${\tt g}$ are constants.   Tr denotes a trace,  indices  $\mu,\nu,\lambda,...$ are raised and lowered  with the Lorentz metric $\eta_{\mu\nu}=$diag$(-,+,+)$, and the totally antisymmetric  symbol $\epsilon_{\mu\nu\lambda}$ is defined such that $\epsilon_{012}=1$.    Extremizing the action with respect to variations in  $X^\mu$ leads to the classical equations of motion
\be [ [X_\mu,X_\nu],X^\nu]+ia\, \epsilon_{\mu\nu\lambda}[X^\nu,X^\lambda] =0\label{eqofmot} \ee 
 The equations of motion (\ref{eqofmot}) are  invariant under:

\noindent i)
 unitary `gauge' transformations, $X^\mu\rightarrow UX^\mu U^\dagger$, where $U$ is an infinite dimensional unitary matrix, 

\noindent
 ii)
 $2+1$ Lorentz transformations $X^\mu\rightarrow L^\mu_{\;\;\nu} X^\nu$,  where $L$ is a $3\times 3$ Lorentz matrix, and 

\noindent iii) 
  translations in the three-dimensional Minkowski space $X^\mu\rightarrow X^\mu+v^\mu\BI$, where $\BI$ is the  unit matrix.

Well known solutions to these equations are non-commutative     $(A)dS^2$\cite{Ho:2000fy}-\cite{Pinzul:2017wch}  and  the   non-commutative  cylinder.\cite{Chaichian:2000ia},\cite{Balachandran:2004yh},\cite{Stern:2014uea}   Both are associated with unitary irreducible representations of three-dimensional Lie algebras. Non-commutative     $(A)dS^2$  corresponds to  unitary irreducible representations  of $su(1,1)$, while the non-commutative  cylinder corresponds to  unitary irreducible representations  of the two-dimensional Euclidean algebra $E_2$.  Thus the former solution is defined by
\be   [X_\mu,X_\nu]=ia\,\epsilon_{\mu\nu\lambda} X^\lambda\qquad\quad X_\mu X^\mu \;\;\,{\rm fixed}\;,\label{ncAdStoo}\ee
while the latter is
\be [X_0,X_\pm] =\pm 2a\,X_\pm\quad\qquad [X_+,X_-]=0\quad\qquad X_+X_-\;\;\,{\rm fixed}\;,\label{tncSxR}\ee
where   $X_\pm=X_1\pm i X_2$.
Non-commutative     $(A)dS^2$  preserves the  Lorentz symmetry ii) of the equations of motion, while the non-commutative  cylinder breaks the symmetry to the two-dimensional rotation group.  Non-commutative     $AdS^2$ was recently shown to be asymptotically commutative, and the holographic principle was applied to map a scalar field theory on non-commutative     $AdS^2$ to a conformal theory on the boundary.\cite{Pinzul:2017wch}

The commutative (or equivalently, continuum or semi-classical) limit  for these two solutions is clearly $a\rightarrow 0$.  Thus  $a$ plays the role of $\hbar$ of quantum mechanics, and for convenience  we shall  make the identification $a=\hbar$ and then take the limit   $\hbar\rightarrow 0$.   In the limit, functions of the matrices  $X_\mu$ are replaced by functions of commutative coordinates $x_\mu$, and to lowest order  in $\hbar$, commutators of functions of $X_\mu$ are replaced by $i\hbar$ times Poisson brackets  of the corresponding functions of $x_\mu$, $\;[{\cal F}(X),{\cal G}(X)]\rightarrow i\hbar \{{\cal F}(x),{\cal G}(x)\}$. 

 So in the  commutative limit of the non-commutative     $(A)dS^2$
solution, (\ref{ncAdStoo}) defines a two-dimensional hyperboloid  with an  $su(1,1)$ Poisson algebra 
\be
\{x_\mu,x_\nu\}=\epsilon_{\mu\nu\lambda}x^\lambda\label{theso21pbs}\ee
Two different geometries result from the two choices of sign for the Casimir in (\ref{ncAdStoo}).  The positive  sign is associated with non-commutative  $(A)dS^2$, while the negative sign  is associated with non-commutative Euclidean $(A)dS^2$.   We describe them below:
\begin{enumerate}
\item Non-commutative $(A)dS^2$.  A positive Casimir yields the constraint $x_\mu x^\mu=r^2$ in the commutative limit,  which defines two-dimensional de Sitter (or anti-de Sitter) space, $(A)dS^2$ (or $H^{1,1}$).   $r$ in this semi-classical solution, and the ones that follow, denotes a constant length scale,  $r>0$.   A global parametrization for   $(A)dS^2$ is given by
\be \pmatrix{x^0\cr x^1\cr x^2}=r \pmatrix{\sinh\tau\cr \cosh\tau\cos\sigma\cr \cosh\tau\sin\sigma}\;,\label{undfrmdads}\ee
where  $-\infty <\tau<\infty$, $0\le\sigma<2\pi$.
Using this parametrization we obtain the following Lorentzian induced metric on the surface:\footnote{$AdS^2$ and $dS^2$ are distinguished by the definition of the time-like direction on the manifold.  For the former, the time parameter corresponds to $\sigma$,  and for the latter, it is $\tau$ .}
\be ds^2=r^2 \, (-d\tau^2+\cosh^2\tau \,d\sigma^2) \ee
The Poisson brackets (\ref{theso21pbs}) are recovered upon writing
\be \{\tau,\sigma\} =\frac 1 {r\,\cosh\tau}\label{tsgmuh}\ee

\item
Non-commutative Euclidean $(A)dS^2$.  A negative Casimir yields the constraint $x_\mu x^\mu=-r^2$ in the commutative limit. This defines a two-sheeted hyperboloid corresponding  to the Euclidean version of  de Sitter (or anti-de Sitter) space, Euclidean  $(A)dS^2$ (or $H^{2,0}$).    A parametrization of   the upper hyperboloid ($x^0>0$)  is
\be \pmatrix{x^0\cr  x^1\cr  x^2}=r \pmatrix{\cosh\tau\cr \sinh\tau\cos\sigma\cr \sinh\tau\sin\sigma}\;,\label{tldtowsgmaprtrs}\ee
where again $-\infty <\tau<\infty$, $0\le\sigma<2\pi$.
Now the  induced metric on the surface has a Euclidean signature
 \be ds^2=r^2 \, (d\tau^2+\sinh^2\tau \,d\sigma^2) \ee
Upon assigning  the Poisson brackets
\be \{\tau,\sigma\} =\frac 1 {r\,\sinh\tau}\label{tsgmuht00}\ee
we again recover the $su(1,1)$ Poisson bracket algebra (\ref{theso21pbs}).  
\end{enumerate}

The  commutative limit of the non-commutative   cylinder
solution (\ref{tncSxR}) is obviously the cylinder.  The Casimir for  the two-dimensional Euclidean algebra goes to $(x^1)^2+(x^2)^2=r^2$, while the limiting Poisson brackets are  $\{x^0,x^1\}=-2{x^2}$,    $\{x^0,x^2\}=2{x^1}$,    $\{x^1,x^2\}=0$.  A parametrization in terms of polar coordinates  
\be\pmatrix{x^0\cr x^1\cr x^2}=\pmatrix{\tau\cr r\cos\sigma\cr r\sin\sigma}
\ee
yields the Lorentzian induced metric
\be  ds^2=-d\tau^2+r^2d\sigma^2\;,\ee
and the Poisson algebra is recovered for $\{\tau,\sigma\}=2 $.

The  above solutions admit either a Euclidean or Lorentzian signature for the induced metric after taking the commutative limit.  The signature for any of these particular solution is fixed.  Below we show that  the inclusion of a mass term in the action allows for solutions with signature change.

\section{Inclusion of a mass term in the $3d$ matrix model}
\setcounter{equation}{0}

We next add a quadratic contribution to   the three-dimensional  Lorentzian matrix model  action   (\ref{mmactn3d}):
\be S(X)=\frac 1{{\tt g}^2}{\rm Tr}\Bigl(-\frac 14 [X_\mu, X_\nu] [X^\mu,X^\nu] +\frac i3  a\, \epsilon_{\mu\nu\lambda}X^\mu[ X^\nu, X^\lambda]+  \frac b2\, X^\mu X_\mu\Bigr)\;\label{mmactnwthmss}\ee
As stated in the introduction, quadratic terms have been shown to result from an IR regularization.\cite{Kim:2011cr}
The  equations of motion resulting from variations of $X^\mu$ are now
 \be [[X_\mu,X_\nu],X^\nu]+i a\epsilon_{\mu\nu\lambda}[X^\nu,X^\lambda]+b X_\mu=0\;\label{flmtrxeqs2d}\ee
As in this article we shall only be concerned with solutions in the commutative limit, $\hbar\rightarrow 0$, we may as well take the the limit of these equations.  In order for the cubic and quadratic  terms to contribute in the commutative limit we need that $a$ and $b$ vanish in the limit according to
\be  a\rightarrow \hbar \alpha \qquad\quad b\rightarrow \hbar^2 \beta\;,\label{clmtfandb}\ee
where  $\alpha$ and $ \beta$ are nonvanishing  and finite.
  The equations (\ref{flmtrxeqs2d}) reduce to
\be  -\{\{x_\mu,x_\nu\},x^\nu\}-\alpha\epsilon_{\mu\nu\lambda}\{x^\nu,x^\lambda\}+\beta x_\mu=0\;\label{Todsceom}\ee

The $AdS^2$ and  Euclidean $AdS^2$ solutions, which are associated with the  $su(1,1)$ Poisson algebra (\ref{theso21pbs}), survive when the mass term is included provided that the constants  $\alpha$ and $ \beta$ are constrained by
\be\beta=2(1-\alpha )\label{betomal}\ee
 In the limit  where the  mass term vanishes,  $\beta=0$ and $\alpha=1$, we recover the solutions of the previous section.  On the other hand, the non-commutative cylinder only solves the equations in the limit of zero mass $\beta\rightarrow 0$.

The mass term allows for new solutions, which have no  $\beta\rightarrow 0$ limit.    One such solution is the fuzzy sphere embedded  in the three-dimensional {\it Lorentzian } background, which was examined in \cite{Chaney:2015mfa}.  In the commutative limit it is defined by
\beqa  &&(x^0)^2+(x^1)^2+(x^2)^2=r^2\cr &&\cr
&&\{x^0,x^1\}=x^2\qquad\{x^1,x^2\}=x^0\qquad\{x^2,x^0\}=x^1\label{fzysphr}\eeqa
These Poisson brackets solve the Lorentzian equations (\ref{Todsceom}) provided that $\alpha=-\frac 12$ and $\beta=-1$.   The solution obviously does not preserve  the  Lorentz symmetry ii) of the equations of motion.
 One can introduce a spherical coordinate parametrization
\be \pmatrix{x^0\cr x^1\cr x^2}=r\pmatrix{\cos\theta\cr\sin\theta\cos\phi\cr \sin\theta\sin\phi}\,
\ee
$0\le \phi<2\pi$, $0< \theta< \pi$.
Then the Poisson brackets in  (\ref{fzysphr}) are recovered for 
$\{\theta,\phi\}=\frac 1r \csc\theta $.  The induced invariant length which one computes from the Lorentzian background,  $ds^2=dx^\mu dx_\mu$, does not give the usual metric for a sphere.  Instead one finds
\be  ds^2=r^2\Bigl(\cos 2\theta \,d\theta^2+\sin^2\theta \,d\phi^2\Bigr)\ee
In addition to the coordinate singularities at the poles,
there are  singularities in the metric at the latitudes $\theta=\frac \pi 4$ and $\frac{3 \pi} 4$.  The Ricci scalar is divergent at these latitudes.    The metric tensor has a Euclidean signature for $ 0< \theta<\frac\pi 4\;\;{\rm and} \;\;\frac {3\pi}4<\theta< \pi$, and a Lorentzian signature for $\frac\pi 4<\theta<\frac {3\pi}4$.  The regions are illustrated in figure 1. The Lorentzian regions of the fuzzy sphere solutions have both an initial and a final singularity and  crudely describe a two-dimensional closed cosmology.  The singularities are resolved away from the commutative limit, where the fuzzy sphere  is expressed in terms of $N\times N$ hermitean matrices.  
Axially symmetric deformations of the fuzzy sphere are also solutions to the Lorentzian matrix model.\cite{Chaney:2015mfa}

Other sets of solutions to the Lorentzian matrix model which have no  $\beta\rightarrow 0$ limit
are deformations of the non-commutative $AdS^2$ and  Euclidean $AdS^2$ solutions.  Like the fuzzy sphere solution, they  break the  Lorentz symmetry ii) of the equations of motion, but preserve spatial rotational invariance.  Again, we shall only be concerned with the commutative limit of these solutions.
\begin{enumerate}
\item Deformed  non-commutative $AdS^2$. 
Here we replace (\ref{undfrmdads}) by
\be \pmatrix{x^0\cr x^1\cr x^2}=r \pmatrix{\sinh\tau\cr \rho\cosh\tau\cos\sigma\cr \rho\cosh\tau\sin\sigma}\;\label{dfrmdads}\ee
 $\rho>0$ is the deformation parameter.  We again assume the  Poisson bracket (\ref{tsgmuh}) between $\tau$ and $\sigma$. 
 Substituting (\ref{dfrmdads}) into  (\ref{Todsceom}) gives
 $\beta=2\rho^2(1-\alpha)=1+\rho^2-2\alpha$.  It is solved by the previous undeformed $AdS^2$ solution, $\rho^2=1$ with (\ref{betomal}), along with  new solutions which allow for arbitrary $\rho>0$, provided that 
\be \alpha=\frac 12 \qquad \beta=\rho^2 \label{al12betr2}\ee
Using the parametrization (\ref{dfrmdads}), the induced invariant interval on the surface is now
\be ds^2=r^2\cosh^2\tau\,\Bigl( (-1+\rho^2\tanh^2\tau)\,d\tau^2+\rho^2\,d\sigma^2\Bigr) \ee
For $\rho^2>1$ the induced metric tensor possesses  space-time singularities at $\tau=\tau_\pm=\pm\tanh^{-1}|\frac 1\rho|$, which are  associated with two signature changes.
For $\tau>\tau_+$ and $\tau<\tau_-$ the signature of the  induced metric is Euclidean, while for $\tau_-<\tau<\tau_+$  the signature of the  induced metric is Lorentzian.  Figure 2 is a plot of deformed $AdS_2$ in the three-dimensional embedding space for  $r=1$, $\rho=1.15$.

\item Deformed  non-commutative  Euclidean $AdS^2$.  We now deform the  upper hyperboloid given in (\ref{tldtowsgmaprtrs}) to
\be \pmatrix{x^0\cr  x^1\cr  x^2}=r \pmatrix{\cosh\tau\cr \rho \sinh\tau\cos\sigma\cr \rho\sinh\tau\sin\sigma}\;,\label{Oneften}\ee
while retaining the Poisson bracket (\ref{tsgmuht00}) between $\tau$ and $\sigma$.
 $\rho$ again denotes the deformation parameter. 
(\ref{Oneften})  with $\rho\ne 0$ is a solution  to (\ref{Todsceom}) provided that the relations (\ref{al12betr2}) again hold.
The induced invariant interval on the surface is now
\be ds^2=r^2\sinh^2\tau\,\Bigl( (\rho^2\coth^2\tau-1)\,d\tau^2+\rho^2\,d\sigma^2\Bigr) \ee
For $\rho^2<1$ there is a singularity at $\tau=\tau_+=\tanh^{-1}|\rho|$ which is associated with a signature change.
For $\tau<\tau_+$  the signature of the induced metric is Euclidean, while for $\tau >\tau_+$ the signature of  the  induced metric is Lorentzian.  Figure 3 gives a plot of deformed hyperboloid in the three-dimensional embedding space for  $r=1$, $\rho=.85$.  The deformed Euclidean $AdS^2$ solution has only an initial (big bang) singularity that appears in the commutative limit, and so, crudely speaking, the Lorentzian region describes an open two-dimensional cosmology.  The singularity is resolved away from the commutative limit.  
\end{enumerate}
\begin{figure}[p]
\centering
\includegraphics[height=2in,width=2.5in,angle=0]{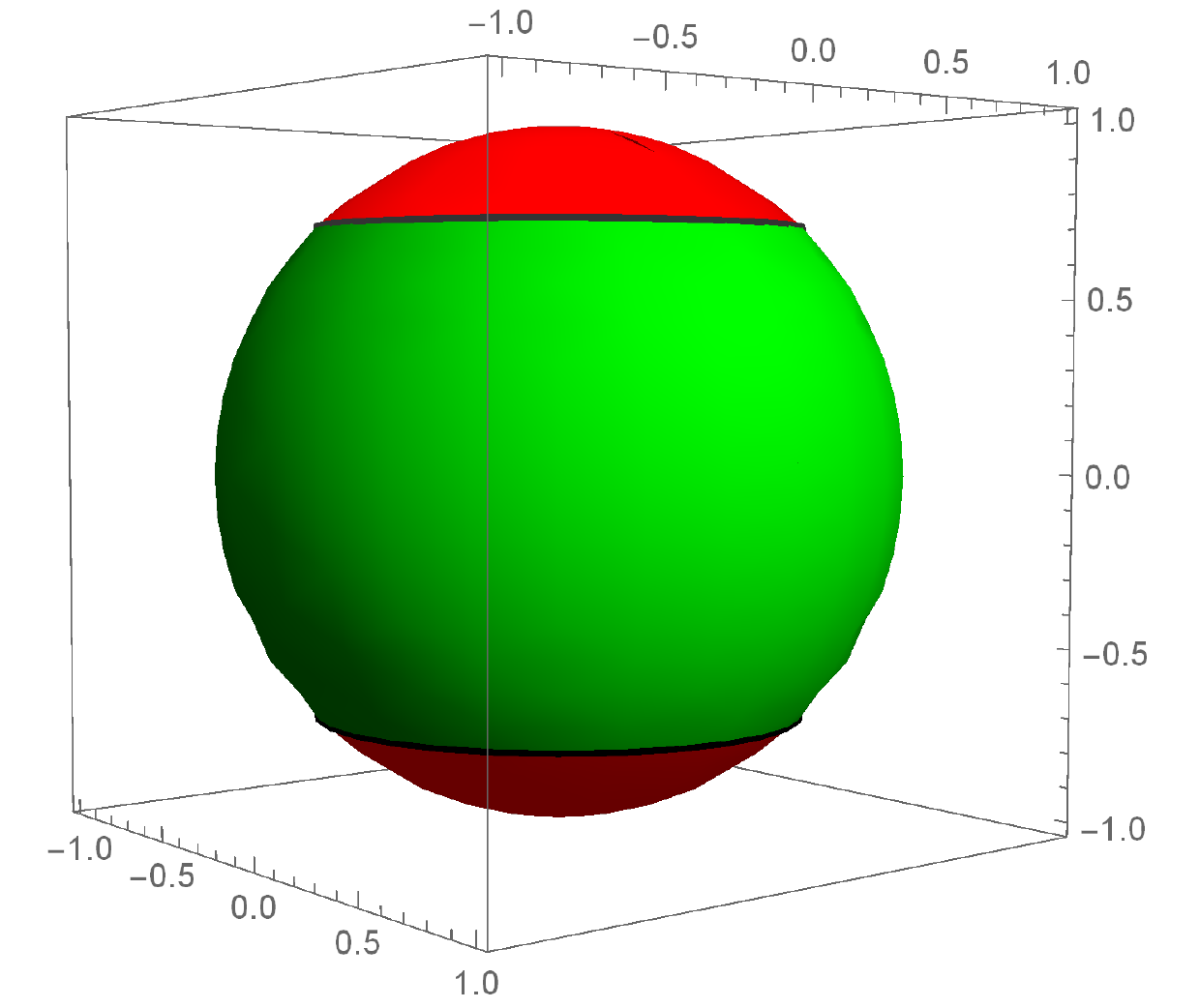}
\caption {Commutative limit of the Lorentzian fuzzy sphere.  Singularities in the metric appear at  $\theta=\frac \pi 4$ and $\frac{3 \pi} 4$, and the signature of the metric changes at these latitudes.  These latitudes are associated with singularities in the Ricci scalar.   The metric tensor has a Euclidean signature for $ 0< \theta<\frac\pi 4\;\;{\rm and} \;\;\frac {3\pi}4<\theta< \pi$ (red regions), and Lorentzian signature for $\frac\pi 4<\theta<\frac {3\pi}4$ (green region).  
}
\includegraphics[height=2in,width=2.5in,angle=0]{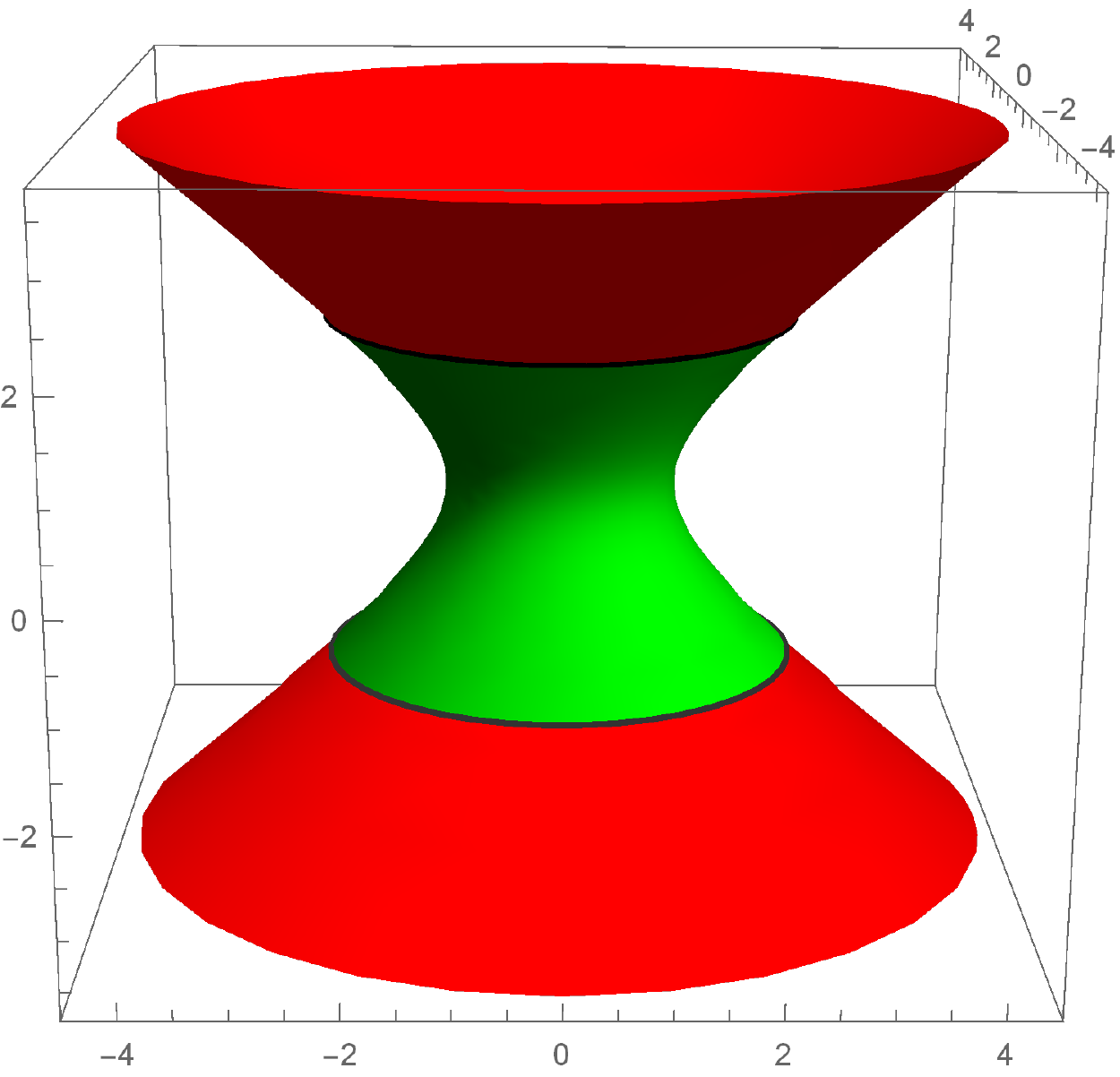}
\caption {Deformed $ AdS_2$ solution with $r=1$, $\rho=1.15$.  The space-time singularities  occur at $\tau=\tau_\pm=\pm\tanh^{-1}|\frac 1\rho|$.  The green region has Lorentzian signature and the red region has Euclidean signature.
}
\includegraphics[height=2in,width=2.5in,angle=0]{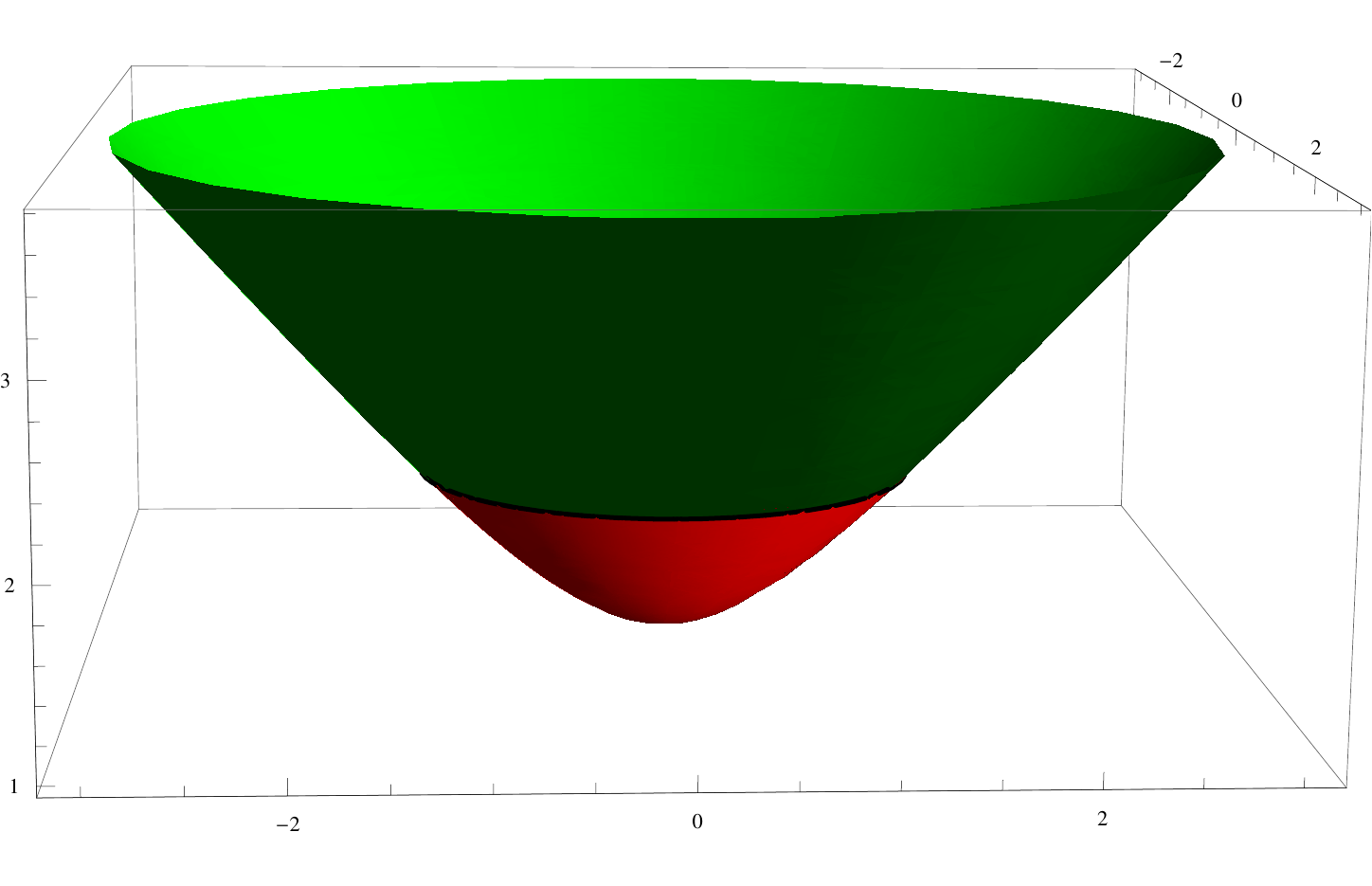}
\caption {Deformed Euclidean $AdS^2$ solution with $r=1$, $\rho=.85$.  A space-time singularity  occurs at $\tau=\tau_+=\tanh^{-1}|\rho|$.  The green region has Lorentzian signature and the red region has Euclidean signature.
}
\end{figure}

\section{ $CP^{1,1}$  and $CP^{0,2}$ solutions}
\setcounter{equation}{0}

Concerning the generalization to four dimensions, a natural approach would be to  examine non-commutative $CP^2$.\cite{Nair:1998bp}-\cite{Balachandran:2005ew}  Actually, if we  wish to recover  noncompact  manifolds in the commutative limit we should consider the indefinite versions  of non-commutative $CP^2$;  non-commutative $CP^{1,1}$ and $CP^{0,2}$.    In this section  we show that non-commutative $CP^{1,1}$ and $CP^{0,2}$ are solutions to an eight-dimensional matrix model with an indefinite background metric.  As with our earlier result,  we find no signature change in the absence of a mass term in the action.  A mass term will be included in the following section. Here, we begin with some general properties of non-commutative $CP^{1,1}$  and $CP^{0,2}$ in the semi-classical limit, and then construct an eight-dimensional matrix model for which they are solutions.

\subsection{Properties}

Non-commutative $CP^{p,q}$ was studied in \cite{Hasebe:2012mz}.
Here we shall only be interested in its semi-classical limit.
 $CP^{p,q}$ are    hyperboloids $H^{2q,2p+1}$ mod $S^1$.  They can be   defined   in terms of $p+q+1$ complex embedding coordinates $z_i,\;i=1,...,p+q+1$, satisfying the $H^{2q,2p+1}$ constraint
\be \sum_{i=1}^{p+1}z_i^*z_i -\sum_{i=p+2}^{p+q+1}z_i^*z_i=1\;,\label{h2q2pp1}\ee along with the identification 
\be z_i \sim e^{i\beta} z_i\label{identfctn}\ee 
$CP^{p,q}$  can   equivalently be  defined  as the coset space $SU(p+1,q)/U(p,q)$.  For the semi-classical limit of non-commutative $CP^{p,q}$ we must also introduce a compatible Poisson structure.  For this we take
\be \{z_i, z^*_
j \} = \left\{ \begin{array}{ll}
     -i\delta_{ij}\;, \;    & \mbox{if $\;i,j=1,...,p+1 $};\\
   \;\;\;   i\delta_{ij}\;,\; & \mbox{if $\;i,j=p+2,...,p+q+1$}\end{array} \right. \;,\label{gnrlpbsfrzzstr}\ee  
while all other Poisson brackets amongst $z_i$ and $z^*_i$ vanish.   Then one can regard (\ref{h2q2pp1}) as the first class constraint that generates the phase equivalence (\ref{identfctn}).

In specializing to $CP^{1,1}$  and $CP^{0,2}$,
 it is convenient to introduce the metric $\eta^C=$diag$(1,1,-1)$ on the three-dimensional complex space spanned by $z_i,\;i=1,2,3$.  Then writing $z^i=(\eta^C)^{ij}z_j$, the constraint  (\ref{h2q2pp1}) for  $CP^{1,1}$ becomes
\be  z^iz_i^*=1\;, \label{cponeone}\ee 
while for  $CP^{0,2}$ the constraint  can be written as
\be \;z^iz_i^*=-1\label{cp0two}\;\ee  
For both cases, the Poisson brackets (\ref{gnrlpbsfrzzstr}) become
\be \{z^i, z^*_
j \} =-i\delta^i_{j}\qquad\;\; \{z^i, z^
j \} =  \{z^*_i, z^*_
j \} = 0\;\label{pbsfrzzstr}\ee

  $CP^{1,1}$ and $CP^{0,2}$ can also be described in terms of  orbits on $SU(2,1)$.   Below  we review  some properties of the Lie algebra $su(2,1)$.
One can write down the defining representation  for  $su(2,1)$ in terms of traceless $3\times 3$ matrices, $\tilde\lambda_a$, $a=1,2,...,8$, which are analogous to the  Gell-Mann matrices $\lambda_a$ spanning $su(3)$.    We denote   matrix elements  by $[\tilde\lambda_a]^i_{\;\,j},\;i,j,..=1,2,3$.  Unlike  $su(3)$ Gell-Mann matrices, $\tilde\lambda_a$  are not all  hermitean, but instead,   satisfy 
\be \tilde\lambda_a\eta^C=\eta^C \tilde\lambda_a^\dagger\label{zroptone4}\ee  They are given in terms of the standard Gell-Mann matrices  in Appendix A. The commutation relations for  $\tilde\lambda_a$ are 
\be [\tilde\lambda_a,\tilde\lambda_b]=2i\tilde f_{abc}\tilde\lambda^c\;,\label{fcrfsu21}\ee
where indices $a,b,c...$ are raised and lowered using the  Cartan-Killing metric on the eight-dimensional space
\be\eta={\rm diag}(1,1,1,-1,-1,-1,-1,1)\,\label{atedeemtrc}\ee 
   $\tilde f_{abc}$ for  $su(2,1)$ are  totally antisymmetric.  Their values, along with some properties of $su(2,1)$, are given in Appendix A.

$CP^{0,2}$ is the coset space $SU(2,1)/U(2)$.  Using the conventions of Appendix A, it is spanned by adjoint orbits in $su(2,1)$ through
 $\tilde\lambda_8$, and consists of elements  $g\tilde\lambda_8g^{-1},\; g\in SU(2,1)$.  The  little group  of  $\tilde\lambda_8$ is  $U(2)$, which is generated by $\tilde\lambda_1,\tilde\lambda_2,\tilde\lambda_3,\tilde\lambda_8$.
On the other hand,   $CP^{1,1}$ is the coset space $SU(2,1)/U(1,1)$.  It is corresponds to orbits through \be\tilde\Lambda_8=\frac 1{\sqrt{3}}\pmatrix{-2&&\cr&1&\cr&&1}= -\frac {\sqrt{3}}2\tilde\lambda_3-\frac {1}2\tilde\lambda_8\;,\ee    $CP^{1,1}=\{g\tilde\Lambda_8g^{-1},\; g\in SU(2,1)\}$.  The little group  of  $\tilde\Lambda_8$ is  $U(1,1)$, which is generated by $\tilde\Lambda_8,\tilde\lambda_6,\tilde\lambda_7,\tilde\Lambda_3=\pmatrix{\;&&\cr&1&\cr&&-1}= -\frac {1}2\tilde\lambda_3+\frac {\sqrt{3}}2\tilde\lambda_8$.

Next, we can construct eight  real coordinates $x^a$ from $z^i$ and $z^*_i$ using
 \be  x^a=z^*_i[\tilde\lambda^a]^i_{\;\,j} z^
j\;,\label{22six}\ee
They are invariant under the phase transformation (\ref{identfctn}), and span a four dimensional manifold.  Using (\ref{frzidntee}), the constraints on the coordinates are
\be x^a x_a=\frac 43\qquad \quad \tilde d_{abc}x^b x^c= \pm \frac 23 x_a\;,\label{cpcnstrnts}\ee 
where one takes the upper sign in the second equation for $CP^{1,1}$  and the lower sign for $CP^{0,2}$. $\tilde d_{abc}$  is totally symmetric; the values are given in Appendix A.
 From (\ref{pbsfrzzstr}), $x_a$ satisfy an $su(2,1)$ Poisson bracket algebra
\be\{x_a,x_b\}=2\tilde f_{abc}x^c\label{cpmtrc}\ee 

\subsection{Eight-dimensional matrix model}

It is now easy to construct an eight-dimensional `IKKT'-type  matrix model for which (\ref{cpmtrc}) is a solution, at least in the commutative limit.  As before we only consider the 
 bosonic sector, spanned by eight infinite-dimensional hermitean matrices $X_a$, with indices raised and lowered with the indefinite flat metric $\eta_{ab}$.   In analogy with the three-dimensional  model in (\ref{mmactn3d}), take the action to  consist of a quartic  term and a cubic term:
\be S(X)=\frac 1{{\tt g}^2}{\rm Tr}\Bigl(-\frac 14 [X_a, X_b] [X^a,X^b] +\frac i3  a\, \tilde f_{abc}X^a[ X^b, X^c]\Bigr)\;\label{mmactn}\ee  The cubic term appears  ad hoc, and we remark that it is actually unnecessary for the purpose of finding  solutions when a quadratic term is introduced instead.  We  consider quadratic terms in  section five.  On the other hand, the cubic term leads to a richer structure for the space of solutions and it is for that reason we shall consider it.

The   equations of motion following from (\ref{mmactn}) are
\be [ [X_a,X_b],X^b]+ia\, \tilde f_{abc}[X^b,X^c] =0\label{CPeqofmot} \ee
They are invariant under 
 unitary `gauge' transformations, $SU(2,1)$ transformations and
  translations. Assuming that the constant $a$  behaves as in (\ref{clmtfandb}) in the  commutative limit, leads to
  \be  -\{\{x_a,x_b\},x^b\}-\alpha\tilde f_{abc}\{x^b,x^c\}=0\;\label{CPTodsceonmt}\ee
The Poisson brackets (\ref{cpmtrc}) solve these equations for $\alpha=2$. They describe a $CP^{1,1}$ or   $CP^{0,2}$ solution, the choice depending on the sign in
the second constraint in (\ref{cpcnstrnts}).

 For either  solution, we can project the eight-dimensional flat metric $\eta$ down to the   surface   $\bar z z=z_i^*z^i=\pm 1\;$, in order to obtain  the induced metric. Once again, $i=1,2,3$. Using  the Fierz identity (\ref{frzidntee}), we get
\be ds^2=dx^a dx_a=4\Bigl((\bar z z)(d\bar z dz)-|\bar zdz|^2\Bigr)\;,\label{FSmtrc}\ee
where $\bar zdz= z^*_
i dz^i$, $d\bar z dz =dz_i^*dz^i$ and we have used $d(\bar z z)=0$. (\ref{FSmtrc}) is the Fubini-Study metric written on a noncompact space.

 We next examine the induced metric tensor  on a local coordinate patch.  We choose the  local coordinates $(\zeta_1,\zeta_2)$, defined by
 \be \zeta_1=\frac {z^1}{z^3}\qquad\quad  \zeta_2=\frac {z^2}{z^3}\;,\qquad z^3\ne 0\;\label{lclcrdnts}\;,\ee
along with their complex conjugates.
 These coordinates  respect the equivalence relation (\ref{identfctn}).  In the language of constrained Hamilton formalism, they are first class variables. From their definition  it follows that $|\zeta_1|^2+|\zeta_2|^2-1=\pm |z^3|^{-2}$ and  $\bar zdz=|z^3|^2\Bigl(\zeta_1^*d\zeta_1+\zeta_2^*d\zeta_2\Bigr)\pm d\log z^3$, where the upper [lower] sign applies for  $CP^{1,1}$ [$CP^{0,2}$].  Substituting into (\ref{FSmtrc}) gives the induced metric tensor on the coordinate patch
\be\frac 14 ds^2\;=\;\frac 12 g_{{}_{\zeta_u \zeta_v^*}} d\zeta_u d\zeta_v^*\;=\;\frac{|d\zeta_1|^2+|d\zeta_2|^2}{|\zeta_1|^2+|\zeta_2|^2-1}\;-\;\frac{|\zeta_1^*d\zeta_1+\zeta_2^*d\zeta_2|^2}{(|\zeta_1|^2+|\zeta_2|^2-1)^2}\;\ee
It has the same form 
for both   $CP^{1,1}$ and $CP^{0,2}$. Because $|\zeta_1|^2+|\zeta_2|^2-1<0$ for the latter, $CP^{0,2}$ has  Euclidean signature. 
 The Poisson brackets  (\ref{pbsfrzzstr}) can  be projected down to the local coordinate patch as well.  The result is
 \beqa \{\zeta_u,\zeta_v^*\}&=&\pm i (|\zeta_1|^2+|\zeta_2|^2-1)(\zeta_u\zeta_v^*-\delta_{uv}) \cr&&\cr\{\zeta_u,\zeta_v\}&=&\{\zeta_u^*,\zeta_v^*\}\;=\;0\;,\qquad u,v=1,2\label{pbmatrx}\eeqa
Once again, the upper [lower] sign applies for  $CP^{1,1}$ [$CP^{0,2}$].
 The resulting symplectic two-form is K\"ahler:
\be \Omega=\mp\frac i2  g_{{}_{\zeta_u \zeta_v^*}} d\zeta_u\wedge d\zeta_v^*\label{omayguh}\ee

Next we re-write the induced metric and symplectic two-form using three Euler-like angles $(\theta,\phi,\psi)$,   $0\le\theta<\pi$,  $0\le\phi<2\pi$,  $0\le \psi<4\pi$, along with one  real variable $\tau$, $-\infty<\tau<\infty$.   We  treat  $CP^{1,1}$ and $CP^{0,2}$ separately:
 \begin{enumerate}
 \item $CP^{1,1}\,.\;$  Now write
 \be \zeta_1=  e^{i(\psi+\phi)/2}\coth\tau\,\cos\frac\theta 2\qquad\qquad\zeta_2=  e^{i(\psi-\phi)/2}\coth\tau\,\sin\frac\theta 2\;,\label{crds11}\ee
 which is consistent with the requirement that $|\zeta_1|^2+|\zeta_2|^2-1>0$. The  induced metic in these coordinates has the Taub-NUT form (which was also true for the $CP^2$ solution\cite{Chaney:2016npa}) 
 \beqa {ds^2}&=&g_{\tau\tau}\,{d\tau^2}+ g_{\theta\theta} \,(d\theta^2+\sin^2\theta d\phi^2)+ g_{\psi\psi}\, (d\psi+\cos\theta d\phi)^2
\label{gnrlfrm}\;\eeqa
We get
 \be g_{\tau\tau}=-4\qquad \quad g_{\theta\theta} =\cosh^2\tau\qquad \quad  g_{\psi\psi}=-\cosh^2\tau\sinh^2\tau\label{tothree2}\;,\ee
with the other nonvanishing components of the induced metric being $ g_{\phi\phi}= g_{\psi\psi}\,\cos^2\theta+ g_{\theta\theta} \,\sin^2\theta$ and  $ g_{\psi\phi}= g_{\psi\psi}\,\cos\theta$.
 The result  indicates that there are two space-like directions and two time-like directions.   The symplectic two-form in terms of these coordinates is 
\beqa \Omega_{{}_{CP^{1,1}}}&=&-\sinh\tau\cosh\tau \,d\tau\wedge (d\psi+\cos\theta\, d\phi)+\frac 12 \cosh^2\tau \sin\theta \, d\theta\wedge d\phi\cr&&\cr&&=-\frac 12 \,d\,\Bigl(\cosh^2\tau \,(d\psi+\cos\theta\, d\phi) \Bigr)\label{smplctoneone}\eeqa

 \item  $CP^{0,2}\,.\;$  Here choose   
 \be \zeta_1=  e^{i(\psi+\phi)/2}\tanh\tau\,\cos\frac\theta 2\qquad\qquad\zeta_2=  e^{i(\psi-\phi)/2}\tanh\tau\,\sin\frac\theta 2\;,\label{crds02}\ee
 which is consistent with the inequality $|\zeta_1|^2+|\zeta_2|^2-1<0$. The resulting induced metric again has the Taub-NUT form (\ref{gnrlfrm}).   In contrasting with (\ref{tothree2}), results differ for the  $g_{\theta\theta}$ component
  \be  g_{\tau\tau}=-4\qquad \quad g_{\theta\theta} =-\sinh^2\tau\qquad \quad  g_{\psi\psi}=-\cosh^2\tau\sinh^2\tau\label{tothree4}\;,\ee
where again  $ g_{\phi\phi}= g_{\psi\psi}\,\cos^2\theta+ g_{\theta\theta} \,\sin^2\theta$ and  $ g_{\psi\phi}= g_{\psi\psi}\,\cos\theta$.
 The induced metric now has a   Euclidean signature, and the symplectic two-form is 
\beqa \Omega_{{}_{CP^{0,2}}}&=&-\sinh\tau\cosh\tau \,d\tau\wedge (d\psi+\cos\theta\, d\phi)+\frac 12 \sinh^2\tau \sin\theta \,d\theta\wedge d\phi \cr&&\cr&&=-\frac 12 \,d\,\Bigl(\sinh^2\tau \,(d\psi+\cos\theta\, d\phi) \Bigr)\label{smplct2zero}\eeqa
 \end{enumerate}  

Both metric tensors (\ref{tothree2}) and (\ref{tothree4}) [including the corresponding results for $ g_{\phi\phi}$ and  $ g_{\psi\phi}$] describing $CP^{1,1}$, and $CP^{0,2}$, respectively, are solutions to the sourceless Einstein equations with cosmological constant $\Lambda=\frac 32$.\footnote{ $CP^2$ is also a solution to the sourceless Einstein equations with cosmological constant $\Lambda=\frac 32$.\cite{Chaney:2016npa} }        Obviously, the metric tensors don't exhibit signature change.    In both cases, the sign of the determinant of the  metric tensor, det$\,g= g_{\tau\tau}\,g_{\psi\psi}\,(g_{\theta\theta}\,\sin\theta)^2$, is positive (away from coordinate singularities).  

 The above discussion utilized the induced metric tensor.  However,  the  relevant metric in the semi-classical limit for a matrix model solution is not, in general, the induced metric, but rather, it is the metric  that appears in the coupling to matter.\cite{Steinacker:2010rh}  This is the so-called `effective' metric tensor, which we here denote by  $\gamma_{\mu\nu}$.  It can be determined from the induced metric  $g_{\mu\nu}$ and the symplectric matrix $\Theta^{\mu\nu}$ using
\be \sqrt{|\det\gamma|}\,\gamma^{\mu \nu}=\frac 1{ \sqrt{|\det\Theta|}}\,[\Theta^Tg\Theta]^{\mu \nu}\,\label{dfefmtrix}\ee
It follows that $|\det\gamma|=|\det g|$, and we can use this identification to determine the effective metric from the induced metric.  We review a derivation of (\ref{dfefmtrix}) in Appendix  B.
 In two dimensions, it is known that  the effective metric is identical to the induced metric, $\gamma_{\mu\nu}=g_{\mu\nu}$.\cite{Arnlind:2012cx}
 This is also  the case for the  $CP^{1,1}$ and  $CP^{0,2}$ solutions, as is shown in Appendix  B, and so all the previous results that followed from the induced metric also apply for the effective metric.   On the other hand, for the solutions of the next section,  in addition to finding signature change, we find that  the effective metric and induced metric for any particular emergent  manifold are in general distinct. 

\section{Inclusion of a mass term in the $8d$ matrix model}
\setcounter{equation}{0}

In analogy to section three, we now add a mass term to the matrix model action (\ref{mmactn}),
\be S(X)=\frac 1{{\tt g}^2}{\rm Tr}\Bigl(-\frac 14 [X_a, X_b] [X^a,X^b] +\frac i3  a\, \tilde f_{abc}X^a[ X^b, X^c]+6\tilde b\, X_aX^a\Bigr)\;\label{mmactnwqt}\ee  
The matrix equations of motion become
 \be [[X_a,X_b],X^b]+i a\tilde f_{abc}[X^b,X^c]+12\tilde b\, X_a=0\;,\label{messu21wm}\ee
 In the semi-classical limit $\hbar \rightarrow 0$, we take 
$a\rightarrow \hbar\alpha$, along with $\tilde b\rightarrow \hbar^2 \tilde\beta$.  Then (\ref{messu21wm}) goes to
\be -\{\{x_a,x_b\},x^b\}- \alpha\tilde f_{abc}\{x^b,x^c\}+12\tilde\beta x_a=0\;,\label{smclmtrxeq}\ee 

These equations are solved by (\ref{cpmtrc}) for
\be \alpha=2(1+\tilde\beta)\label{btaalpm2}\ee
Thus  $CP^{1,1}$ and $CP^{0,2}$ are solutions to the massive matrix model.  In the limit where the  mass term vanishes,  $\tilde \beta=0$ and $\alpha=2$, we recover the solutions of the previous section. $CP^{1,1}$ and $CP^{0,2}$ solutions also  persist in the absence of the  cubic term in the matrix model action (\ref{mmactnwqt}).
For this we need  $\alpha=0$ and $\tilde\beta=-1$.
The mass term allows for other solutions which have no  $\beta\rightarrow 0$ limit. 
Among these solutions are the deformations of  $CP^{1,1}$ and  $CP^{0,2}$, as well as deformations of $CP^2$, which we discuss in the following subsections.

\subsection{Deformations of  $CP^{1,1}$ and  $CP^{0,2}$}
 
For  deformations of  $CP^{1,1}$ and  $CP^{0,2}$ we modify the ansatz  (\ref{22six}) to
\beqa  x_{1-3}&=&\mu\, z^*_i[\tilde\lambda_{1-3}]^i_{\;\,j} z^
j\cr &&\cr x_{4-7}&=&\;\; z^*_i[\tilde\lambda_{4-7}]^i_{\;\,j} z^
j\cr &&\cr x_{8}&=&\nu\, z^*_i[\tilde\lambda_{8}]^i_{\;\,j} z^
j\;,\label{dfrmdnstz}\eeqa
where $\mu$ and $\nu$ are deformation parameters, which we shall restrict to  be real.
 This is a solution to the equations (\ref{smclmtrxeq}) provided that the following relations hold amongst the parameters:
\beqa (2\mu-\alpha)\Bigl(\mu^2+\frac 12\Bigr)+3 \mu\tilde\beta&=&0\cr&&\cr
 \mu^2+\nu^2+2-\alpha(\mu+\nu)+4\tilde\beta&=&0\cr&&\cr
 2\nu-\alpha+2\nu \tilde\beta&=&0\label{cndtnsnmnab}\eeqa
 These relations reduce to (\ref{btaalpm2}) when $\mu=\nu=1$, and so we recover  undeformed $CP^{1,1}$ and  $CP^{0,2}$ in this limit.  For generic values of the parameters,
there are nontrivial solutions to these algebraic relations, which can be expressed as functions of the mass parameter $\tilde\beta$.  For a particular choice of signs:
 \be \alpha ={2 \mu} \,\frac{   \tilde\beta^2-\tilde\beta  -1- \gamma[\tilde\beta]}{ 2  \tilde\beta+1}\qquad\quad\mu={\sqrt{\frac{\tilde\beta^3 -4\tilde\beta^2 -6\tilde\beta + \tilde\beta\gamma[\tilde\beta]-2}{2(  \tilde\beta^2 +4\tilde\beta+2)}}}\qquad\quad\nu=\frac \alpha{2(1+\tilde\beta)}\;,\label{cmplctedsln}\ee
 where \be \gamma[\tilde\beta]=\sqrt{\tilde\beta^4-12 \tilde\beta^3-22 \tilde\beta^2-12\tilde\beta-2 }\label{gmma}\ee
Upon requiring $\gamma[\tilde\beta]$ to be real, we obtain  three disconnected intervals $i-iii)$ in $\tilde\beta$:
 \beqa i)\;\;\qquad\qquad\qquad\qquad\qquad\qquad\qquad\qquad\qquad\quad &\tilde\beta&\le \;\,\frac{1}{2} \left(6-3 \sqrt{6}-\sqrt{98+40 \sqrt{6}}\right)\approx -0.746\cr &&\cr ii)\quad \frac{1}{2} \left(6-3 \sqrt{6}+\sqrt{98+40 \sqrt{6}}\right)\approx -0.603\;\, \le&\tilde\beta&\le \,\;\frac{1}{2} \left(6+3 \sqrt{6}-\sqrt{98+40 \sqrt{6}}\right)\approx -0.325\cr&&\cr iii)\quad \frac{1}{2} \left(6+3 \sqrt{6}+\sqrt{98+40 \sqrt{6}}\right)\;\approx\;\, 13.67\;\, \le
&\tilde\beta&\;\eeqa
We   further restrict $\mu$ to be  real. (Reality of  $\alpha$ and $\nu$ then follows.)  For solution (\ref{cmplctedsln}), this reduces the acceptable regions in $\tilde\beta$ to
\be i')\;-3.414\lesssim\tilde\beta\lesssim -0.746\;,\qquad\quad  \;ii')\;\; -0.603\,\lesssim\,\tilde\beta\, \lesssim\, -0.586\;,\qquad \quad\;iii) \;13.67\,  \lesssim\,
\tilde\beta\ee
  When $\tilde\beta=-.6\, $,  we recover the undeformed case $\mu=\nu=1$ (along with $\alpha=.8)\,$.  Therefore, matrix  solutions in the range $\;ii')$ can be regarded as continuous deformations of the undeformed solutions, while those in the ranges $\;i')$ and  $\;iii)$ cannot be continuously connected to the undeformed solutions.

In addition to the family of solutions given in (\ref{cmplctedsln}) and (\ref{gmma}), the equations (\ref{cndtnsnmnab}) have the simple solution:
\be \alpha=\nu=0\qquad \quad \tilde \beta=-\frac 3 5\qquad \quad \mu^2=\frac 2 5\label{spclsln}\ee
It is a solution for the case where the  cubic term in the matrix model action (\ref{mmactnwqt}) is absent.  From (\ref{dfrmdnstz}), $\nu=0$ implies that the projection of the solution along the  $8^{\rm th}$-direction vanishes, $x_8=0$.  This solution is not contained in (\ref{cmplctedsln}) and (\ref{gmma}).

The ansatz (\ref{dfrmdnstz})  for $\mu$ and $\nu$ not both equal to one, leads to two  types of solutions: 
\begin{enumerate}
\item
Deformed  $CP^{1,1}$, where the complex coordinates $z^i$ satisfy the constraint   (\ref{cponeone}), and  
\item Deformed $CP^{0,2}$, where the complex coordinates $z^i$ satisfy   (\ref{cp0two}).  
\end{enumerate}
 We next  compute the induced metric for these two  types of solutions.

\subsubsection{Induced metric}

The induced metric is again computed   by projecting the eight-dimensional flat  metric (\ref{atedeemtrc}) onto the surface.  From the ansatz (\ref{dfrmdnstz}) we get
\beqa ds^2=dx^a dx_a&=&4(|z_1|^2+|z_2|^2-|z_3|^2)(|dz_1|^2+|dz_2|^2-|dz_3|^2)-4| z^*_1dz_1+z^*_2dz_2-z^*_3dz_3|^2\cr&&\cr&+&4(\mu^2-1)(|z_1|^2+|z_2|^2)(|dz_1|^2+|dz_2|^2)+(\mu^2-1)(z^*_1dz_1+z^*_2dz_2-z_1dz^*_1-z_2dz^*_2)^2\cr&&\cr&+&\frac 13(\nu^2-1)( z^*_1dz_1+z^*_2dz_2+2z^*_3dz_3+z_1dz^*_1+z_2dz^*_2+2z_3dz^*_3)^2
\cr&&\label{dfrmmtrc}\eeqa
We have not yet specialized to the two cases 1. and 2.

The result  (\ref{dfrmmtrc}) can be rewritten in terms of the  local coordinates $(\zeta_1,\zeta_2)$,  defined in (\ref{lclcrdnts}), according to
\beqa ds^2&=& 4|z_3|^2\Bigl( -|z_3|^{2}|\Xi|^2\pm (|d\zeta_1|^2+|d\zeta_2|^2)\Bigr)\cr&&\cr&+&4(\mu^2-1)(|z_3|^2\pm 1)\Bigl((1\pm |z_3|^{-2})|dz_3|^2+ |z_3|^2(|d\zeta_1|^2+|d\zeta_2|^2)+\Xi \,z_3dz_3^*+\Xi^*z_3^*dz_3\,\Bigr)\cr&&\cr&+&(\mu^2-1)\Bigl( |z_3|^2(\Xi-\Xi^* )+(1\pm |z_3|^{-2})(z^*_3dz_3
-z_3dz_3^*)\Bigr)^2\;+\;(\nu^2-1)\Bigl(d|z_3|^2\Bigr)^2\;,\cr&&
\label{dfmtczta}\eeqa
where $\Xi=\zeta_1^* d\zeta_1+\zeta_2^* d\zeta_2$ and we have  used  $|\zeta_1|^2+|\zeta_2|^2=1\pm |z_3|^{-2}$.  The upper [lower] sign applies for deformed $CP^{1,1}$ [$CP^{0,2}$].  The expression (\ref{dfmtczta}) simplifies after making the gauge choice that $z_3$ is real, which we shall do below. 

 The signature  of the induced  metric becomes more evident after expressing  it in terms of the three Euler-like angles $\theta,\phi,\psi$, along with parameter $\tau$ spanning $R_+$, as we did in section four for the undeformed metrics.  For this we now specialize to the two cases: 1.   deformed $CP^{1,1}$ and  2. deformed $CP^{0,2}$ .
 \begin{enumerate}
 \item  Deformed $CP^{1,1}\,.\;$

For this case we  can  apply coordinate transformation (\ref{crds11}). Upon making the phase choice  $z_3=\sinh\tau$, (\ref{dfmtczta}) can  be written in the Taub-NUT form (\ref{gnrlfrm}), where the metric components are now
\beqa g_{\tau\tau}&=&4\Bigl((\mu^2+\nu^2-2) \cosh^2\tau\,{\sinh^2\tau} -1\Bigr)\cr&&\cr g_{\theta\theta}&=&\cosh^2\tau\,(\mu^2\cosh^2\tau-\sinh^2\tau) \cr&&\cr  g_{\psi\psi}&=&- \cosh^2\tau\sinh^2\tau
\label{tofor2}\eeqa
The remaining nonvanishing components of the induced metric are again  obtained from $ g_{\phi\phi}= g_{\psi\psi}\,\cos^2\theta+ g_{\theta\theta} \,\sin^2\theta$ and  $ g_{\psi\phi}= g_{\psi\psi}\,\cos\theta$.  
  The undeformed  $CP^{1,1}$ induced metric tensor in (\ref{tothree2})  is recovered from
(\ref{tofor2})  upon setting $\mu=\nu=1$. This limit thus corresponds to there being
 two space-like directions and two time-like directions, with sign$(g_{\tau\tau},g_{\theta\theta})=(-,+)$ and det$\;g>0$ (away from coordinate singularities).  The same  two space-like directions and two time-like directions appear  in the  limit $|\tau |\rightarrow 0$. Signature change can occur when we go away from either of these two  limits, as we describe below:  

For the solutions   given by (\ref{cmplctedsln}) and (\ref{gmma}), given some value of $\tilde \beta$ $(\ne -1,-.6)$  in the regions $i'$, $ii')$ and $iii)$, the sign of either $g_{\tau\tau}$ or  $ g_{\theta\theta}$  changes at some value of $|\tau|$.  We plot the values of  $|\tau|$ versus $\tilde \beta$ for which this occurs in figure 4.  $g_{\tau\tau}$ changes   sign when  $(\mu ^2+\nu ^2-2)\sinh^2 \tau\cosh^2\tau=1\,$ (indicated by the red curves in figure 4).  $ g_{\theta\theta}$ changes sign when $\tanh^2\tau=\mu^2$ (indicated by the green curves in figure 4).  Above the red curves,   sign $(g_{\tau\tau},g_{\theta\theta})=(+,+)$ and det$\;g<0$  and so the induced metric has Lorentzian signature in this region.  In this case, the time-like direction corresponds to $d\psi+\cos\theta d\phi$.  It corresponds to a space-time with closed time-like curves.  Above the green curves,  sign$(g_{\tau\tau},g_{\theta\theta})=(-,-)$,  while det$\;g>0$. In this case, the induced metric space has a Euclidean signature. 

For the solution (\ref{spclsln}), a sign change in $g_{\theta\theta}$ occurs at $\tanh^2\tau =\frac 25$, and  the induced metric has a Euclidean signature for $\tanh^2\tau >\frac 25$.
\begin{figure}
\centering
\begin{subfigure}{.325\textwidth}
  \centering
  \includegraphics[height=1.8in,width=1.8in,angle=0]{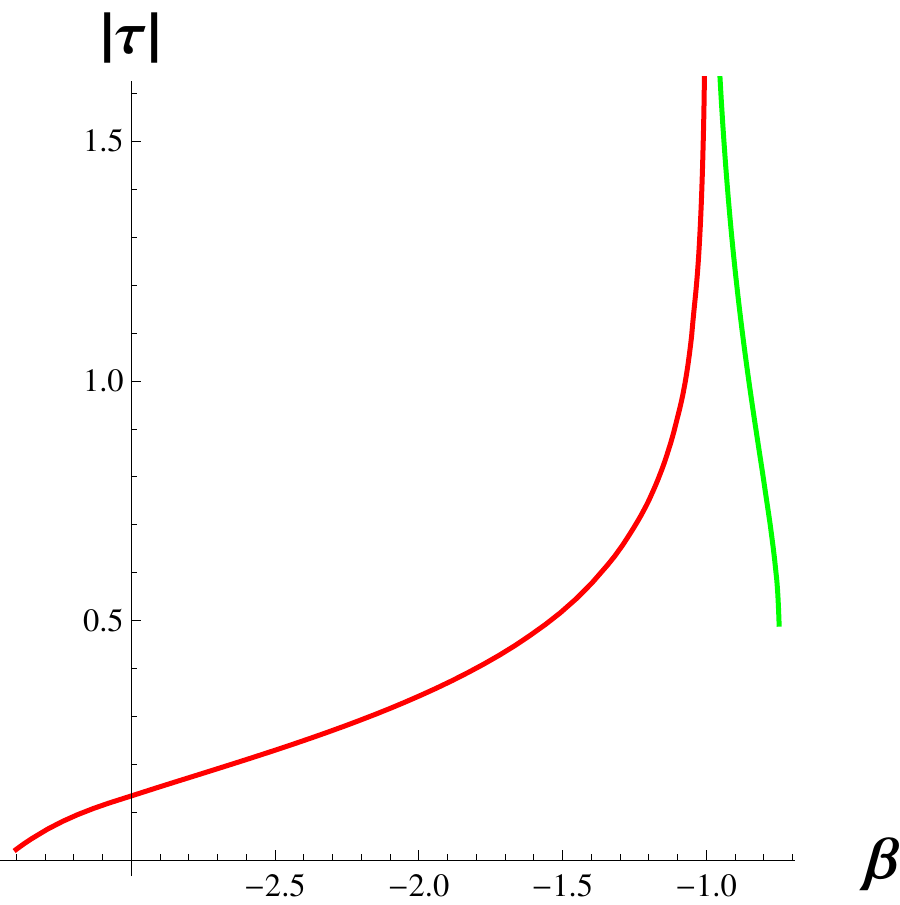}
  \caption{region $i')$}
  \label{fig:sub1}
\end{subfigure}%
\begin{subfigure}{.325\textwidth}
  \centering
  \includegraphics[height=1.8in,width=1.8in,angle=0]{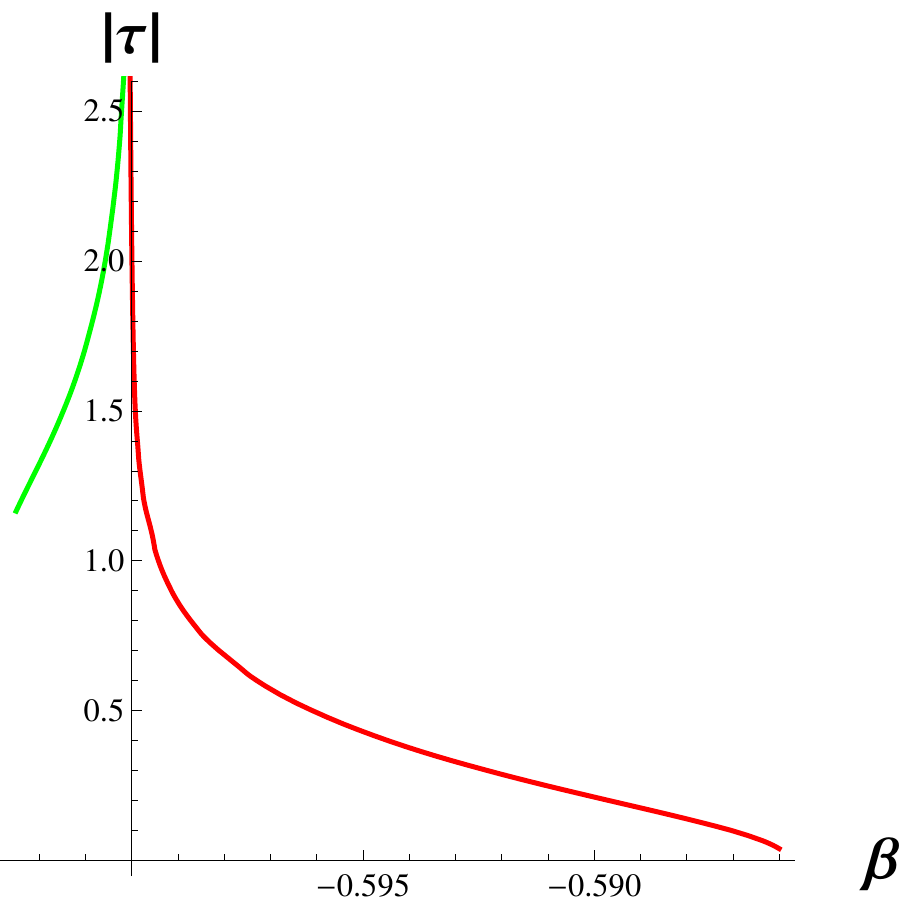}
  \caption{region $\;ii')$}
  \label{fig:sub1}
\end{subfigure}%
\begin{subfigure}{.325\textwidth}
  \centering
  \includegraphics[height=1.8in,width=1.8in]{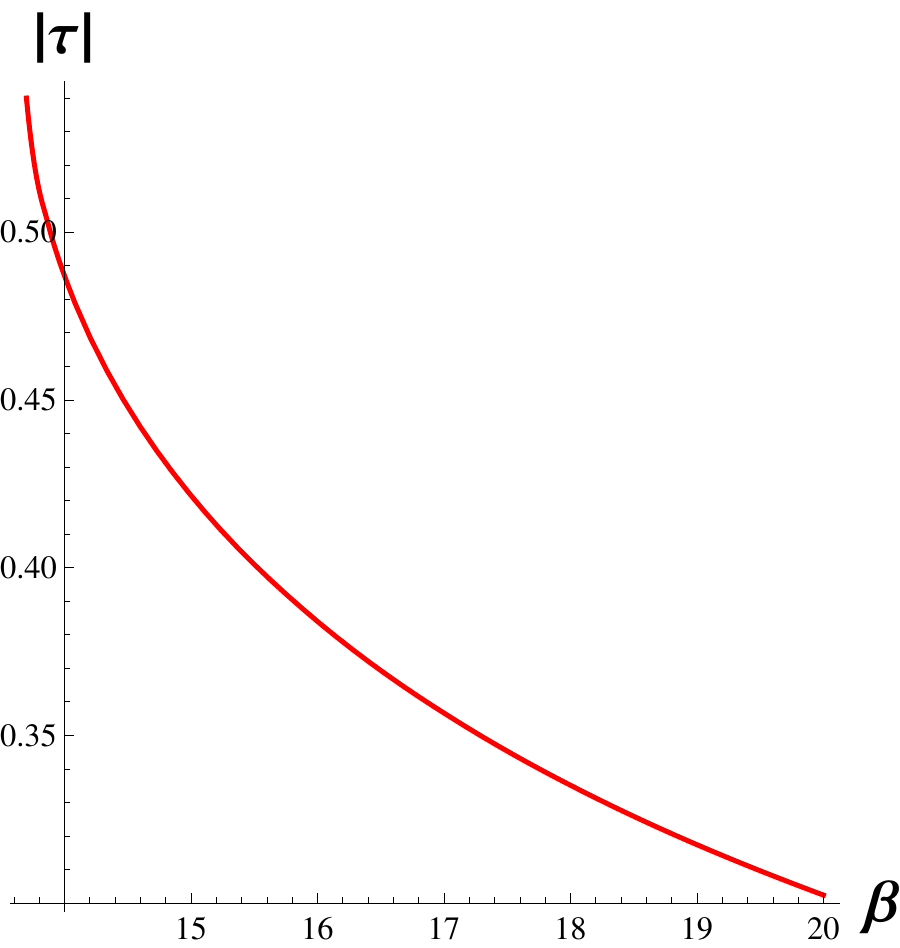}
  \caption{region $iii)$}
  \label{fig:sub2}
\end{subfigure}
\caption{Signature changes in the induced metric  $g_{\mu\nu}$ and effective metric 
 $\gamma_{\mu\nu}$ for   deformed $CP^{1,1}$ are given in plots of $|\tau |$ versus $\tilde\beta$ for  in the three disconnected  regions:  $i')\;-3.414\lesssim\tilde\beta\lesssim -0.746$ (subfigure a), $\;ii')\;\; -0.603\,\lesssim\,\tilde\beta\, \lesssim\, -0.586\;$ (subfigure b) and  $iii)\;13.67\;\lesssim\,\tilde\beta$ (subfigure c).   A sign change in  $g_{\theta\theta}$  or $\gamma_{\theta\theta}$  is indicated by the green curves.
Sign change   in  $g_{\tau\tau}$ or $ \gamma_{\psi\psi}$ is indicated by the red  curves. }
\label{fig:test}
\end{figure}

\item  Deformed $CP^{0,2}\,.\;$

Here  we apply the  coordinate transformation (\ref {crds02}) to (\ref{dfmtczta}), along with the phase choice $z_3=\cosh\tau$.  The induced invariant interval again takes the   Taub-NUT form (\ref{gnrlfrm}), with the matrix elements now being
  \beqa  g_{\tau\tau}&=&4\Bigl((\mu^2+\nu^2-2) \cosh^2\tau\,{\sinh^2\tau} -1\Bigr)\cr&&\cr g_{\theta\theta}&=&  \sinh^2\tau\,\Bigl(\mu^2\sinh^2\tau-\cosh^2\tau \Bigr)\cr&&\cr  g_{\psi\psi}&=&- \cosh^2\tau\sinh^2\tau
\label{tofor3}\;,\eeqa
and  $ g_{\phi\phi}= g_{\psi\psi}\,\cos^2\theta+ g_{\theta\theta} \,\sin^2\theta$ and  $ g_{\psi\phi}= g_{\psi\psi}\,\cos\theta$.
Only the results for $ g_{\theta\theta}$ differ in   expressions (\ref{tofor2}) and (\ref{tofor3}).  The latter reduce to that of undeformed $CP^{0,2}$, (\ref{tothree4}), when $\mu=\nu=1$.  For that limit, as well as for $|\tau |\rightarrow 0$,  sign$(g_{\tau\tau},g_{\theta\theta})=(-,-)$ and det$\;g>0$  (away from coordinate singularities).  In this case, the induced metric has a Euclidean signature.    
As with deformed $CP^{1,1}$, signature change can occur when we go away from  these   limits, as we describe below:  

For the solutions   given by (\ref{cmplctedsln}) and (\ref{gmma}), we find that for any fixed value of  $\tilde\beta$ in the regions  $i')$, $\;ii')$  and  $iii)$, either a sign change occurs for {\it both} $g_{\tau\tau}$ and  $ g_{\theta\theta}$, or there is no signature change.   
We plot the signature changes  for deformed $CP^{0,2}$   in figure 5.   $ g_{\theta\theta}$ changes sign when $\coth^2\tau=\mu^2$ (indicated by the green curves in figure 5).   $g_{\tau\tau}$ changes sign when $(\mu ^2+\nu ^2-2)\sinh^2 \tau\cosh^2\tau=1\,$ (indicated by the red curves in figure 5).   We find that there are no sign changes in the induced metric  for  $-1<\tilde\beta\lesssim -0.746$ and 
 $ -0.603\,\lesssim\,\tilde\beta\,\le -.6$.  So for these sub-regions, the signature of the induced metric remains Euclidean for all $\tau$. For the complementary sub-regions, a sign change  occurs in  $g_{\tau\tau}$, say at $|\tau|=|\tau_1|$,  
and $g_{\theta\theta}$, at a later $|\tau|$, say $|\tau_2|$, i.e.,    $|\tau_2|>|\tau_1|$. For  $|\tau|>|\tau_2|$,   sign$(g_{\tau\tau},g_{\theta\theta})=(+,+)$, and so the induced metric  has a Lorentzian signature in this case.  The time-like direction corresponds to $d\psi+\cos\theta d\phi$, once again corresponding to a space-time with closed time-like curves. For  the intermediate interval in $|\tau|$ where  $|\tau_1|<|\tau|<|\tau_2|$, we get sign$(g_{\tau\tau},g_{\theta\theta})=(+,-)$.  In this case, the induced metric has a  Lorentzian signature, with $\tau$ defining  the time-like direction. Any $\tau-$slice is topologically a three-sphere, since from (\ref {crds02}), 
\be |\zeta_1|^2+ |\zeta_2|^2=\tanh^2\tau\label{towsliceone1}\ee
 Restricting to positive $\tau$,   the  interval  $\tau_1<\tau<\tau_2$ has an initial  singularity at  $\tau_1$,  and final  singularity at  $\tau_2$. Therefore, although  not very realistic, it describes a closed space-time cosmology.

No signature change in the induced metric results from the solution (\ref{spclsln}).
\end{enumerate}
\begin{figure}
\centering
\begin{subfigure}{.325\textwidth}
  \centering
  \includegraphics[height=1.8in,width=1.8in,angle=0]{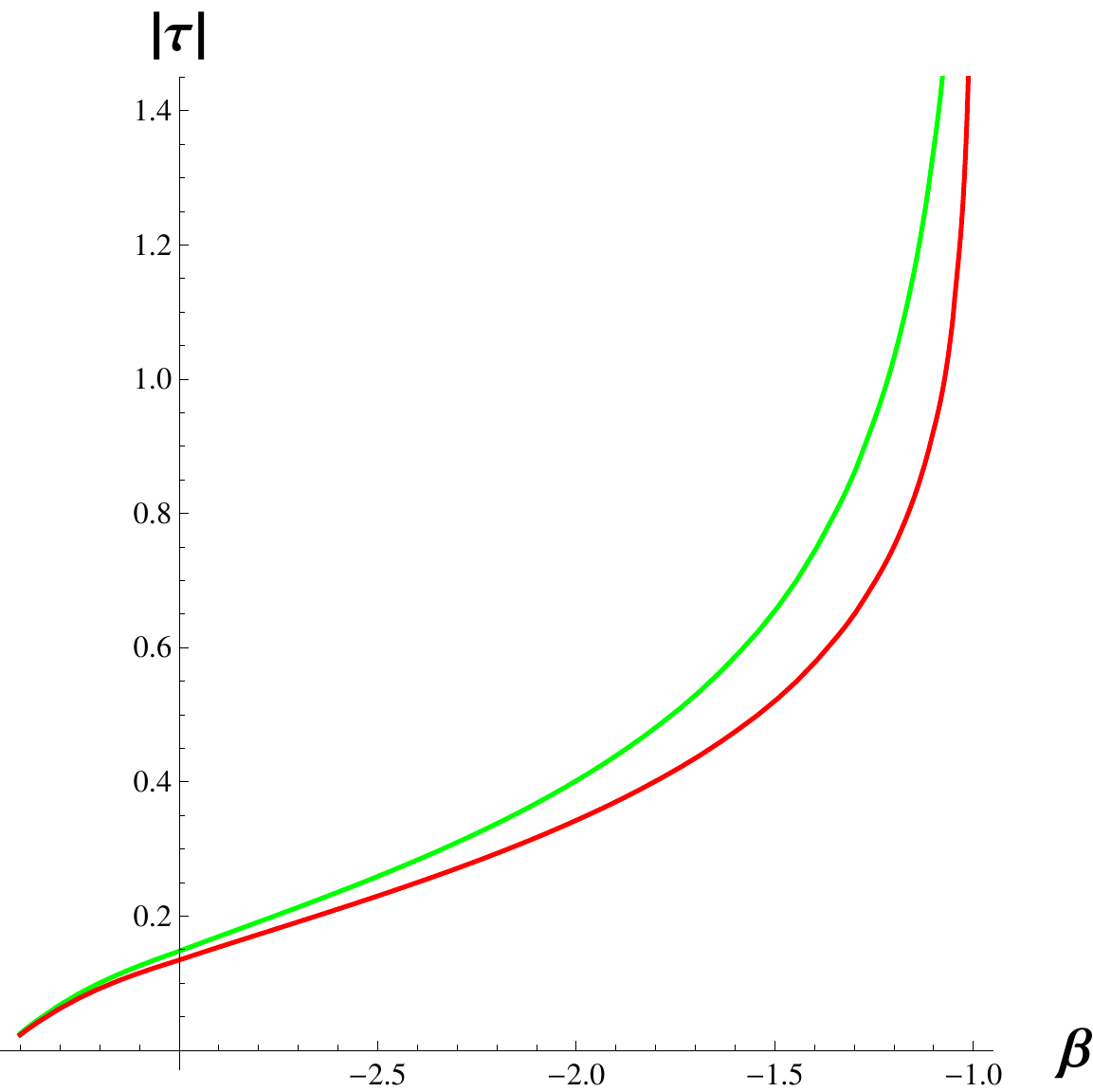}
  \caption{region $i')$}
  \label{fig:sub1}
\end{subfigure}%
\begin{subfigure}{.325\textwidth}
  \centering
  \includegraphics[height=1.8in,width=1.8in,angle=0]{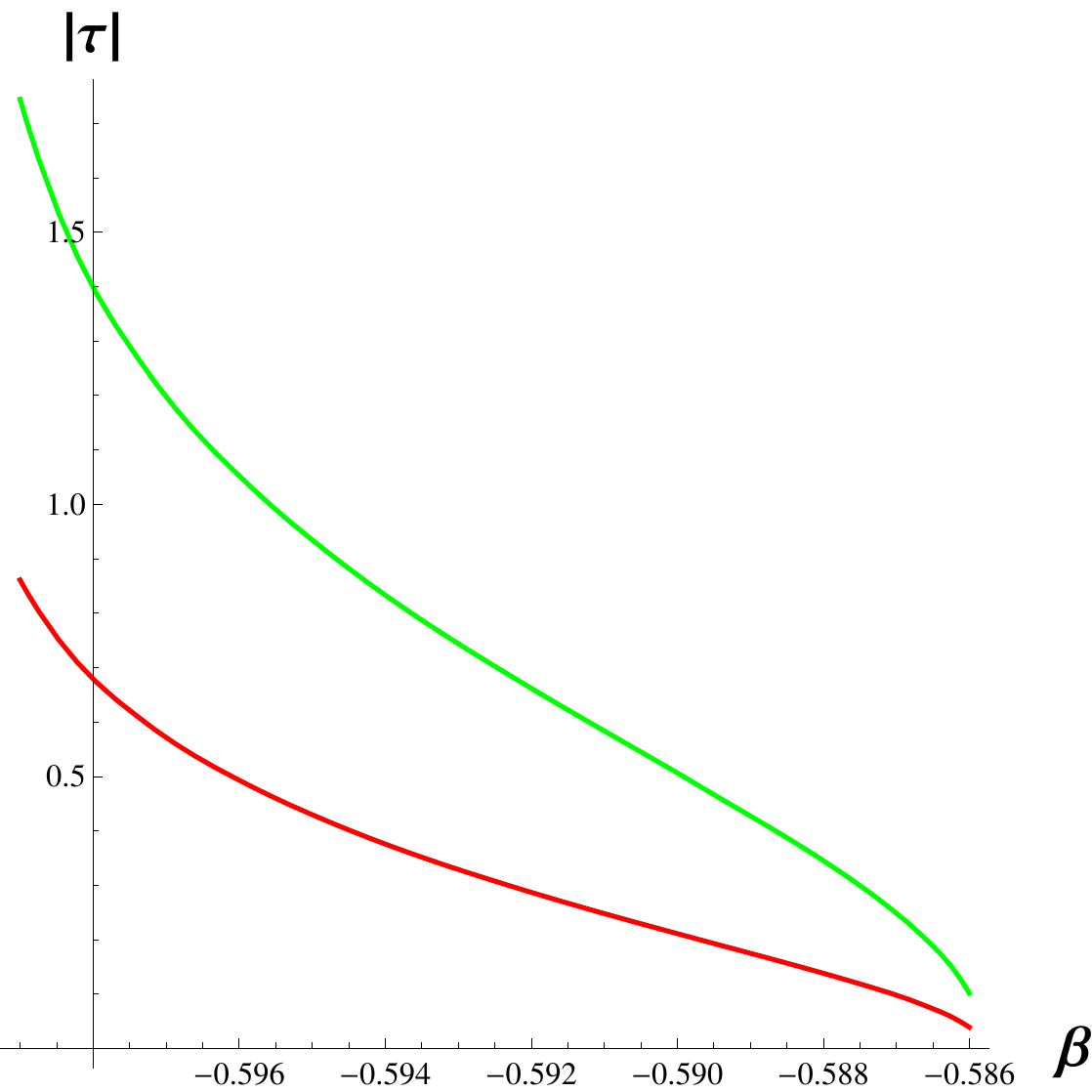}
  \caption{region $\;ii')$}
  \label{fig:sub1}
\end{subfigure}%
\begin{subfigure}{.325\textwidth}
  \centering
  \includegraphics[height=1.8in,width=1.8in]{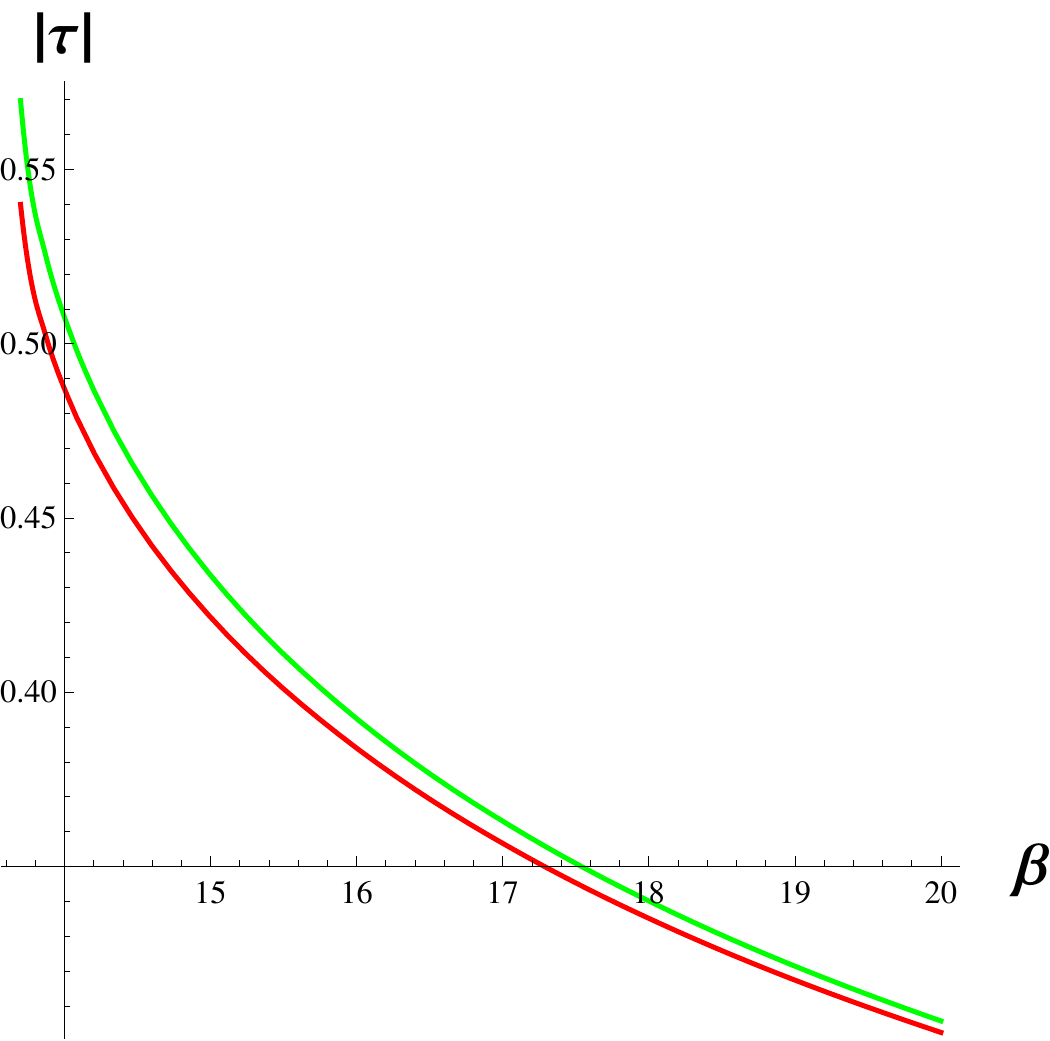}
  \caption{region $iii)$}
  \label{fig:sub2}
\end{subfigure}
\caption{Signature changes in the induced metric  $g_{\mu\nu}$ and effective metric 
 $\gamma_{\mu\nu}$ for   deformed $CP^{0,2}$ are given in plots of $|\tau |$ versus $\tilde\beta$ for  in the three disconnected  regions:  $i')\;-3.414\lesssim\tilde\beta\lesssim -0.746$ (subfigure a), $\;ii')\;\; -0.603\,\lesssim\,\tilde\beta\, \lesssim\, -0.586\;$ (subfigure b) and  $iii)\;13.67\;\lesssim\,\tilde\beta$ (subfigure c).   A sign change in  $g_{\theta\theta}$  or $\gamma_{\theta\theta}$  is indicated by the green curves.
A sign change   in  $g_{\tau\tau}$ or $ \gamma_{\psi\psi}$ is indicated by the red curves.  No sign changes occur in either the induced metric or effective metric for  $-1<\tilde\beta\lesssim -0.746$ and 
 $ -0.603\,\lesssim\,\tilde\beta\,\le -.6\,$.}
\label{fig:test}
\end{figure}

We remark that while the induced metrics for the two solutions 1. and 2. are modified from their undeformed counterparts,  their Poisson brackets, and corresponding symplectic two-forms, are unchanged.  That is, for deformed $CP^{1,1}$ the symplectic two-form is (\ref{smplctoneone}) and  for deformed $CP^{0,2}$ symplectic two-form is (\ref{smplct2zero}).  This is relevant for the computation of the effective metric, which we do in the following subsection.

\subsubsection{Effective metric}

In section four we found that  the induced metric $g_{\mu\nu}$  and effective metric $\gamma_{\mu\nu}$ for undeformed   $CP^{1,1}$ and  $CP^{0,2}$ are identical.
The same result does not hold for the corresponding deformed solutions, as we show below.  Furthermore, more realistic cosmologies follow from the effective metric of the deformed   $CP^{1,1}$ and  $CP^{0,2}$ solutions.

 \begin{enumerate}
 \item  Deformed $CP^{1,1}\,.\;$
To compute the effective metric we need the symplectic matrix, as well as the induced metric.
For the deformed, as well as undeformed, $CP^{1,1}$ solutions, the nonvanishing components of the inverse symplectic matrix are given in (\ref{nvrssmplctc11}), while the induced metric for deformed $CP^{1,1}$ is given by (\ref{tofor2}).  In addition, $|\det\Theta|$ is given in  (\ref{detsfrcp11}), while
$\; |\det\gamma|$  gets deformed, such that
\be |\det\gamma||\det\Theta|= 4|g_{\tau\tau}|\,(\mu^2\cosh^2\tau-\sinh^2\tau)^2\label{dtgmthtcp11}\;,\ee
with $g_{\tau\tau}$ given in (\ref{tofor2}).
As a result, the nonvanishing components of the effective metric tensor are given by
\beqa \frac{ \gamma_{\tau\tau}}{ \sqrt{|\det\gamma||\det\Theta|}}&=&-1\cr&&\cr 
\frac{ \gamma_{\theta\theta}}{ \sqrt{|\det\gamma||\det\Theta|}}&=&\frac 1{4(\mu^2-\tanh^2\tau)}
\cr&&\cr \frac{\gamma_{\psi\psi}}{ \sqrt{|\det\gamma||\det\Theta|}}&=&
-\frac{1}{4({\rm sech}^2\tau\,{\rm csch}^2\tau+2 -\mu ^2-\nu ^2) }\;,\label{fmt11}\eeqa
in addition to  $ \gamma_{\phi\phi}= \gamma_{\psi\psi}\,\cos^2\theta+ \gamma_{\theta\theta} \,\sin^2\theta$ and  $ \gamma_{\psi\phi}= \gamma_{\psi\psi}\,\cos\theta$.
The results again agree with the undeformed induced metric (\ref{tothree2}) in the $\mu=\nu=1$ limit.  This limit has two space-like directions and two time-like directions,  with sign$(\gamma_{\psi\psi},\gamma_{\theta\theta})=(-,+)$ and det$\;\gamma>0$ (away from coordinate singularities).  Signature changes occur  in the effective metric for the same values of the parameters at which the signature changes occur  for the induced metric.

For the solutions   given by (\ref{cmplctedsln}) and (\ref{gmma}), signature changes are again given in figure 4. A sign change  in $\gamma_{\theta\theta}$  [as  with $g_{\theta\theta}$]   appears  when $\tanh^2\tau=\mu^2$  (indicated by the green curves in figure 4).  The  effective metric has a Euclidean signature above the green curves.
A sign change  in  $ \gamma_{\psi\psi}$ [as   with $g_{\tau\tau}$] appears when  $(\mu ^2+\nu ^2-2)\sinh^2 \tau\cosh^2\tau=1\,$ (indicated by the red curves in figure 4).  Above the red curves, the signature of the effective metric  is Lorentzian, det$\;\gamma<0$,  and $\tau$ is the  time-like direction.  A $\tau-$slice again defines a three-sphere, since from (\ref {crds11}), 
\be |\zeta_1|^2+ |\zeta_2|^2=\coth^2\tau\ee
 Restricting to positive $\tau$,   this region with Lorentzian signature has an initial  singularity, and therefore, and it describes an open space-time.  We shall see in the next subsection that it corresponds to an expanding cosmology.

For the solution (\ref{spclsln}), a sign change in $\gamma_{\theta\theta}$ occurs at $\tanh^2\tau =\frac 25$, and  the effective metric has a Euclidean signature for $\tanh^2\tau >\frac 25$.  There are no regions with Lorentzian signature in this case.

\item  Deformed $CP^{0,2}\,.\;$
We repeat the above calculation to get  the effective metric    $\gamma_{\mu\nu}$ for  deformed $CP^{0,2}\,.$   The  inverse symplectic matrix is the same as for undeformed $CP^{0,2}\,,$  with nonvanishing components (\ref{nverssmpzero2}).  The induced metric for  deformed $CP^{0,2}$ is given in (\ref{tofor3}).  Using the result for  $|\det\Theta|$  in  (\ref{detsfrcp02}), we now get
\be |\det\gamma||\det\Theta|= 4|g_{\tau\tau}|\,(\mu^2\sinh^2\tau-\cosh^2\tau)^2\label{2sixtee}\;,\ee
with $g_{\tau\tau}$ given  in  (\ref{tofor3}).
Now  the nonvanishing components of the effective metric tensor are found to be
\beqa \frac{ \gamma_{\tau\tau}}{ \sqrt{|\det\gamma||\det\Theta|}}&=&-1\cr&&\cr 
\frac{ \gamma_{\theta\theta}}{ \sqrt{|\det\gamma||\det\Theta|}}&=&\frac 1{4(\mu^2-{\rm coth}^2\tau)}
\cr&&\cr \frac{\gamma_{\psi\psi}}{ \sqrt{|\det\gamma||\det\Theta|}}&=&
\frac{1 }{4(\mu ^2+\nu ^2-2-4\,{ \rm csch}^2 2 \tau) }\;,\label{fmtzro2}\eeqa
again  with  $ \gamma_{\phi\phi}= \gamma_{\psi\psi}\,\cos^2\theta+ \gamma_{\theta\theta} \,\sin^2\theta$ and  $ \gamma_{\psi\phi}= \gamma_{\psi\psi}\,\cos\theta$.  The results reduce to the undeformed induced metric (\ref{tothree4}) in the limit $\mu=\nu=1$,  describing a space with Euclidean signature.

For the solution   given by (\ref{cmplctedsln}) and (\ref{gmma}),  signature changes in the effective metric occur at the same values of the parameters as the signature changes for the induced metric, which are indicated in figure 5.
 $ \gamma_{\theta\theta}$ [like  $ g_{\theta\theta}$] changes sign when
${\rm coth}^2\tau=\mu^2$  (indicated by the green curves in figure 5).
 $\gamma_{\psi\psi}$ [like $g_{\tau\tau}$] changes sign when $(\mu ^2+\nu ^2-2)\sinh^2 \tau\cosh^2\tau=1\,$  (indicated by the red curves in figure 5).  
 As seen in the figure, given any fixed value of  $\tilde\beta$ in the regions $i')\;, \;ii')$ and $iii)$, either a sign change occurs in  both $\gamma_{\psi\psi}$ and  $ \gamma_{\theta\theta}$, or there is no signature change.
 No sign changes in the effective metric  for  $-1<\tilde\beta\lesssim -0.746$ and 
 $ -0.603\,\lesssim\,\tilde\beta\,\le -.6$.  So for these sub-regions the signature of the induced metric remains Euclidean. 
For the complementary regions, a sign change  occurs in $\gamma_{\psi\psi}$ at $|\tau|=|\tau_1|$ and  $ \gamma_{\theta\theta}$ at  and $|\tau|=|\tau_2|$, with  $|\tau_2|>|\tau_1|$.  In the intermediate region  $|\tau_1|<|\tau|<|\tau_2|$, sign$(\gamma_{\psi\psi},\gamma_{\theta\theta})=(+,-)$.  Here the effective metric has a Lorentzian signature, but unlike what happens with the induced metric,  $d\psi+\cos\theta d\phi$ is associated with the time-like direction, yielding closed time-like curves. 
 For  $|\tau|>|\tau_2|$, i.e., above the green curves,  sign$(\gamma_{\psi\psi},\gamma_{\theta\theta})=(+,+)$, and so the effective metric  picks up a Lorentzian signature, with $\tau$ being the time-like direction.  From (\ref{towsliceone1}), a $\tau-$slice is a 3-sphere. Restricting to positive $\tau$,  this region with  Lorentzian signature has an initial  singularity, and  so describes a open space-time cosmology, which we next show, is expanding.

No signature change in the effective metric results from the solution (\ref{spclsln}).
\end{enumerate}

\subsubsection{Expansion}

From the deformed $CP^{1,1}$ and $CP^{0,2}$ solutions we found regions in parameter space where the effective metric has a Lorentzian signature, and possessed an initial singularity.  Any time ($\tau$) - slice is a three-sphere, or more precisely, a Berger sphere.  These examples correspond to  space-time cosmologies with a big bang.  To show that they are expanding we introduce a spatial scale $a(|\tau|)$.  We define it as the cubed root of the three-volume at any $\tau-$slice
\be a(|\tau|)^3=\int_{S^3}\sqrt{|\det \gamma^{(3)}|}\,d\theta d\phi d\psi\;,\ee
where $\gamma^{(3)}$ denotes the effective metric on the $\tau-$slice.  From the form of the metric tesnor,  det$\,\gamma^{(3)}= \gamma_{\psi\psi}\,(\gamma_{\theta\theta}\,\sin\theta)^2$, and since $\gamma_{\psi\psi}$ and $\gamma_{\theta\theta}$ only depend on $\tau$. Then
\be a(|\tau|)^3=(4\pi )^2 \sqrt{|\gamma_{\psi\psi}|}\,|\gamma_{\theta\theta}|\label{sptlscl}\ee  
We wish to determine how the spatial scale evolves with respect to the proper time $t$ in the co-moving frame
\beqa t(\tau)&=&\int_{\tau_0}^{\tau} \sqrt{-\gamma_{\tau\tau}(\tau')}\;d\tau'\,\cr&&\cr &=&\,\int_{\tau_0}^{\tau}  |\det\gamma(\tau')|^{\frac 14}|\det\Theta(\tau')|^{\frac 14}\,d\tau'\label{comvngtm}\eeqa
The lower integration limit $\tau_0$ corresponds to the value of $\tau$ at the big bang, i.e., the signature change. 

 We next compute and plot $a(|\tau|)$ versus $t(\tau)$ for  the two cases, deformed $CP^{1,1}$ and deformed $CP^{0,2}$, in the regions of Lorentzian signature:

\begin{enumerate}

\item Deformed $CP^{1,1}$.  For the spatial volume, we get
\be a(|\tau|)^3
= {(4\pi)^2\,\cosh^3\tau|\sinh\tau|\,|\mu^2\cosh^2\tau-\sinh^2\tau|^{\frac 12}\, \Big|(\mu^2+\nu^2-2) \cosh^2\tau\,{\sinh^2\tau} -1\Big|^{\frac 14}}\;,\label{afrcp1one}\ee
after substituting (\ref{tofor2}) and (\ref{dtgmthtcp11}) into (\ref{sptlscl}).  For the proper time $t(|\tau|)$ in the co-moving frame we get
\be t(\tau)\,={2}\,\int_{\tau_0}^{\tau}\,{|\mu^2\cosh^2\tau'-\sinh^2\tau'|}^{\frac 12} \;\Big|(\mu^2+\nu^2-2) \cosh^2\tau'\,{\sinh^2\tau'} -1\Big|^{\frac 14 } d\tau'\;,\label{prptmcp1one}\ee and  $\tau_0$ is associated with the signature change, given by $\; {\sinh^22\tau_0} =\frac 4{(\mu^2+\nu^2-2)}$.  It    corresponds to value of $\tau$ at  the initial singularity, where from (\ref{afrcp1one}), the spatial scale vanishes. In figure  6(a) we plot $a(|\tau|)$ versus $t(\tau)$ for regions of deformed $CP^{1,1}$ where the effective metric has Lorentzian signature, using three values of $\tilde \beta$. It shows a very rapid expansion near the origin.  For  $\tau$ close to $\tau_0$,  (\ref{afrcp1one}) and (\ref{prptmcp1one}) give $a\sim (\tau-\tau_0)^\frac 1{12}$ and $t\sim (\tau-\tau_0)^\frac 54$.  Hence, $a\sim t^\frac 1{15}$.  For large $\tau$, $a$ is linear in $t$.  The same large distance behavior was found for solutions in \cite{Steinacker:2017vqw}.

\item
For deformed $CP^{0,2}$, (\ref{sptlscl}) gives
\be a(|\tau|)^3
= { (4\pi) ^2 \, |\sinh\tau|^3\cosh\tau|\mu^2\sinh^2\tau-\cosh^2\tau |^{\frac 12}\Big|(\mu^2+\nu^2-2) \cosh^2\tau\,{\sinh^2\tau} -1\Big|^{\frac 14}}\;, \label{COoh2atow}
\ee  
after using (\ref{tofor3}) and from (\ref{2sixtee}).  (\ref{comvngtm}) gives 
\be t(\tau)\,=2\,\int_{\tau_0}^{\tau} {|\mu^2\sinh^2\tau'-\cosh^2\tau'|}^{\frac 12} \; \Big|(\mu^2+\nu^2-2) \cosh^2\tau'\,{\sinh^2\tau'} -1\Big|^{\frac 14}\,d\tau'\label{prptmcpoh2}\ee
Again, the initial value $\tau_0$ for $\tau$ is associated with a signature change, now satisfying $\coth^2\tau_0=\mu^2$. It corresponds to a  big-bang  singularity, and from (\ref{COoh2atow}),  $a(|\tau_0|)=0$.   In figure  6(b) we plot $a(|\tau|)$ versus $t(\tau)$ for regions of deformed $CP^{0,2}$ where the effective metric has Lorentzian signature, using three values of $\tilde \beta$.  It too shows a  rapid expansion near the origin.  For  $\tau$ close to $\tau_0$,  (\ref{COoh2atow}) and (\ref{prptmcpoh2}) give $a\sim (\tau-\tau_0)^\frac 1{6}$ and $t\sim (\tau-\tau_0)^\frac 32$.  Hence, $a\sim t^\frac 19$.  As with the case of deformed $CP^{1,1}$ and  \cite{Steinacker:2017vqw}, $a\sim t$  for large $\tau$.

\end{enumerate}

\begin{figure}
\centering
\begin{subfigure}{.35\textwidth}
  \centering
  \includegraphics[height=1.8in,width=2.in,angle=0]{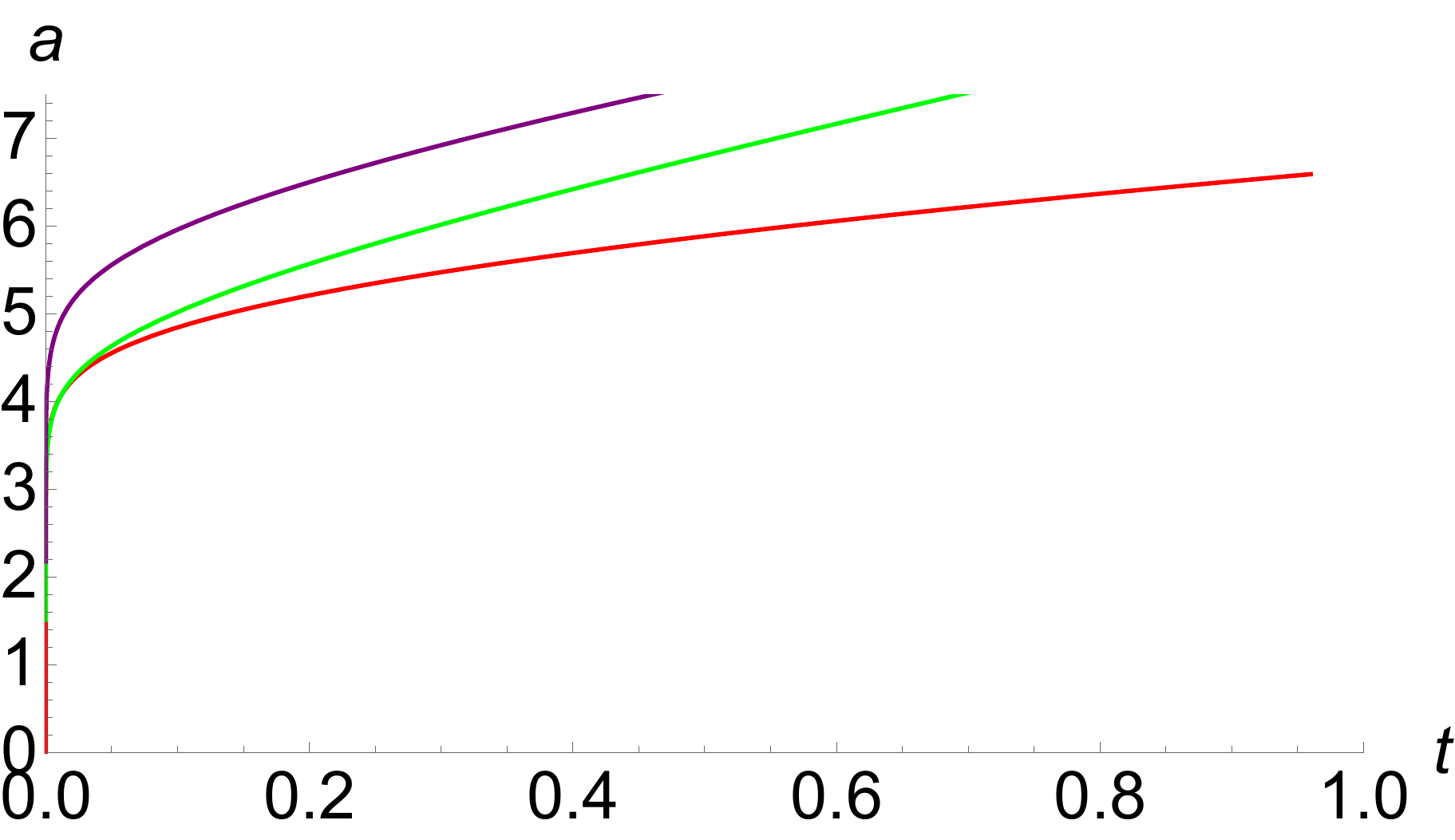}
  \caption{deformed $CP^{1,1}$}
  \label{fig:sub1}
\end{subfigure}%
\begin{subfigure}{.45\textwidth}
  \centering
  \includegraphics[height=1.8in,width=2.in,angle=0]{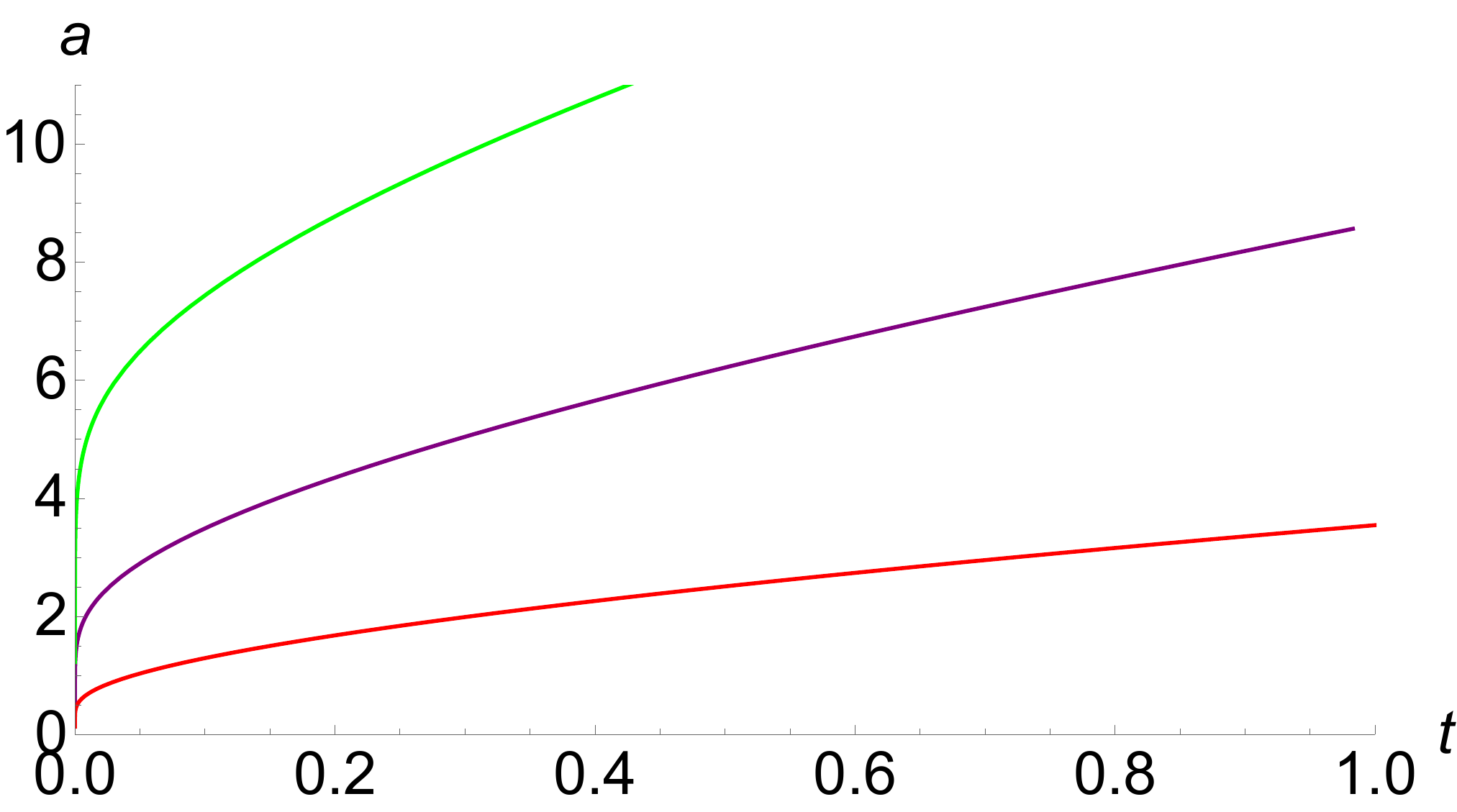}
  \caption{deformed $CP^{0,2}$}
  \label{fig:sub1}
\end{subfigure}%
\caption{ $a(|\tau|)$ versus $t(\tau)$ for regions of deformed $CP^{1,1}$ (subfigure a) and deformed $CP^{0,2}$ (subfigure b) where the effective metric has Lorentzian signature (and $\tau$ is the time-like direction) for  $\tilde \beta=-3$ (red curve), $-.595$ (green curve) and $14$ (purple curve). }
\label{fig:test}
\end{figure}

\subsection{Deformed $CP^2$}

We now look for solutions to  the previous eight-dimensional matrix model which are deformations of non-commutative $CP^2$.  $CP^2$ is a solution to an  eight-dimensional matrix model in a Euclidean background. In \cite{Chaney:2016npa},  such solutions were found when the background metric was changed to diag($+,+,+,+,+,+,+,-)$.  Here we show  that deformations of non-commutative $CP^2$ solve the matrix model with the indefinite metric (\ref{atedeemtrc}), and that they may be associated with multiple signature changes.

We once again assume that the three complex coordinates $z_i$ satisfy the constraint (\ref{cponeone}), but now that the indices are raised and lowered with the three-dimensional {\it Euclidean} metric.
The Poisson brackets that arise from the commutative limit of fuzzy $CP^2$ are (\ref{pbsfrzzstr}) [now, assuming the Euclidean metric] \cite{Chaney:2016npa}.
 We replace the $su(2,1)$ Gell-Mann matrices $\tilde\lambda_a$ in (\ref{dfrmdnstz}) by $su(3)$ Gell-Mann matrices $\lambda_a$, i.e.,
\beqa  x_{1-3}&=&\mu\, z^*_i[\lambda_{1-3}]^i_{\;\,j} z^
j\cr &&\cr x_{4-7}&=&\;\; z^*_i[\lambda_{4-7}]^i_{\;\,j} z^
j\cr &&\cr x_{8}&=&\nu\, z^*_i[\lambda_{8}]^i_{\;\,j} z^
j\;,\label{su3dfrmdnstz}\eeqa
Now substitute this ansatz into the equations of motion (\ref{smclmtrxeq}) to get the following conditions on the parameters
\beqa
(2\mu-\alpha)\Bigl(\mu^2-\frac 12\Bigr)+3 \mu\tilde\beta&=&0\cr&&\cr
\mu^2+\nu^2-2-\alpha(\mu+\nu)+4\tilde\beta&=&0\cr&&\cr
2\nu(\tilde\beta-1)+\alpha&=&0\;,\label{5thirdy}
\eeqa
which differs from  (\ref{cndtnsnmnab}) in various signs.
We can obtain (\ref{5thirdy}) by making the replacement $(\alpha,\tilde\beta,\mu,\nu)\rightarrow (i\alpha,-\tilde\beta,i\mu,i\nu) $ in  (\ref{cndtnsnmnab}).  To obtain a solution to  (\ref{5thirdy}), we can then make the same  replacement in the solution  (\ref{cmplctedsln}).   The result is
 \be \alpha ={2 \mu} \,\frac{   \tilde\beta^2+\tilde\beta  -1- \gamma[-\tilde\beta]}{- 2  \tilde\beta+1}\qquad\quad\mu={\sqrt{\frac{\tilde\beta^3 +4\tilde\beta^2 -6\tilde\beta + \tilde\beta\gamma[-\tilde\beta]+2}{2(  \tilde\beta^2 -4\tilde\beta+2)}}}\qquad\quad\nu=\frac \alpha{2(1-\tilde\beta)}\;,\ee
 where $ \gamma[\tilde\beta]$ was defined in (\ref{gmma}).  The parameters 
 $\mu$, $\nu$, and  $\alpha$ (and necessarily, $  \gamma[-\tilde\beta]$) are all real only for the following two disconnected intervals in $\tilde\beta$:
\be  i) \; .325\,\lesssim\,\tilde\beta\lesssim\,.586\qquad\quad ii) \;3.41\,\lesssim\,\tilde\beta\label{cp2drsfb}\ee

\subsubsection{Induced metric}
The metric induced from the flat  background metric (\ref{atedeemtrc}) onto the surface spanned by (\ref{su3dfrmdnstz}) is
\beqa ds^2=dx^a dx_a&=&-ds^2_{FS}+\Bigl(1+\frac 1{\mu^2}\Bigr)(dx_1^2+dx_2^2+dx_3^2)+\Bigl(1+\frac 1{\nu^2}\Bigr)dx_8^2\;,\label{dssqwfstd}
\eeqa
where $ds^2_{FS}$ denotes the Fubini-Study metric
\beqa ds^2_{FS}=\sum_{a=1}^8\Bigl(d (z^\dagger\lambda_{a}z )\Bigr)^2&=&4(|dz|^2-|z^\dagger dz|^2)
\eeqa
Here we introduce the notation  $|dz|^2=dz^*_i dz^i$, $z^\dagger dz=z^*_i dz^i$ and $z^\dagger\lambda_{a}z=z^*_i[\lambda_{a}]^i_{\;\,j} z^j$.  Then (\ref{dssqwfstd}) becomes
\beqa ds^2&=&-4(|dz|^2-|z^\dagger dz|^2)\cr&&\cr&+&4(\mu^2+1)(|z_1|^2+|z_2|^2)(|dz_1|^2+|dz_2|^2)+(\mu^2+1)(z^*_1dz_1+z^*_2dz_2-z_1dz^*_1-z_2dz^*_2)^2\cr&&\cr&+&\frac 13(\nu^2+1)( z^*_1dz_1+z^*_2dz_2-2z^*_3dz_3+z_1dz^*_1+z_2dz^*_2-2z_3dz^*_3)^2
\eeqa

Next introduce  local coordinates $(\zeta_1,\zeta_2)$ defined in (\ref{lclcrdnts}), now  satisfying $|\zeta_1|^2+|\zeta_2|^2+1=
 |z^3|^{-2}$.  Then
\beqa ds^2&=& -4(z_3)^2\,\Bigl(  |d\zeta_1|^2+|d\zeta_2|^2-(z_3)^{2}|\Xi|^2\Bigr)\qquad
\cr&&\cr&+&4(\mu^2+1)(1-(z_3)^2)\biggl((dz_3)^2\Bigl(\frac 1{(z_3)^2}-1\Bigr)+(z_3)^2(  |d\zeta_1|^2+|d\zeta_2|^2)+\,z_3dz_3(\Xi^*+\Xi)\biggr)\cr&&\cr&+&(\mu^2+1)\biggl((z_3)^2(\Xi-\Xi^*)\biggr)^2\;-\;(\nu^2+1)\Bigl(2z_3dz_3\Bigr)^2\;,\cr&&\label{cp2dfmtczta}\eeqa
where  we again chose $z_3$ to be real, and defined $\Xi=\zeta_1^* d\zeta_1+\zeta_2^* d\zeta_2$.
We introduce  Euler-like angles $(\theta,\phi,\psi)$, along with $\tau$, which now is an angular variable, $0\le\tau<\frac \pi 2$, using
\be \zeta_1= e^{\frac i2(\psi+\phi)}\cos\frac\theta 2\tan\tau\qquad\quad \zeta_2=  e^{\frac i2(\psi-\phi)}\sin\frac\theta 2\tan\tau\label{prtzseepee2}\ee
It then follows that $(z_3)^2=\cos^2\tau$.  
The  induced invariant interval again takes the   Taub-NUT form (\ref{gnrlfrm}), with the non-vanishing matrix elements 
  \beqa  g_{\tau\tau}&=&4\Bigl(-1+(\mu^2-\nu^2)\sin^2\tau\cos^2\tau\Bigr)\cr&&\cr g_{\theta\theta}&=&\sin^2\tau\,(-\cos^2\tau+\mu^2\sin^2\tau) \cr&&\cr  g_{\psi\psi}&=&-\sin^2\tau\cos^2\tau
\label{ndcdfrcptoo}\;,\eeqa
along with  $ g_{\phi\phi}= g_{\psi\psi}\,\cos^2\theta+ g_{\theta\theta} \,\sin^2\theta$ and  $ g_{\psi\phi}= g_{\psi\psi}\,\cos\theta$.
 The induced metric has  Euclidean signature for $\tau$ close to zero.
 A sign change in $ g_{\theta\theta}$ occurs for $\tan\tau=\frac 1{|\mu |}$.  If $\mu^2-\nu^2>\frac 14$, {\it two} additional signature  changes  occur in the induced metric for the domain $0<\tau<\frac \pi 2$. Specifically,  $ g_{\tau\tau}$ changes sign when $\sin 2\tau=\frac 2{\sqrt{\mu^2-\nu^2}}$.  We find numerically, that $\mu^2<\nu^2$ for solutions (\ref{cmplctedsln}) with $\tilde\beta$ in the region $i)$ in (\ref{cp2drsfb}), and  that $\mu^2>\nu^2$ in the region $ ii)$.  So only one signature change occurs when  $\tilde\beta$ has the values in $i)$.   It is a change from the Euclidean signature to one where the induced metric has two space-like directions and two time-like directions. 

 On the other hand, three signature changes can occur when  $\tilde\beta$ has values in  $ii)$.  They are plotted as a function of $\tilde\beta$ in figure seven.   A sign change in  $g_{\theta\theta}$   is indicated by the green curve, and  sign changes   in  $g_{\tau\tau}$ are indicated by the red and blue curves.  The induced metric has  Euclidean signature below the red curve. 
In the tiny intermediate region  between the red and green curves, the induced metric has two-space-like directions and two time-like directions.   It  has Lorentzian signature in the other intermediate region between the green and blue curves, with  $d\psi+\cos\theta d\phi$ time-like. Above the blue curve, the induced metric again has two-space-like directions and two time-like directions.
\begin{figure}
\centering
  \includegraphics[height=1.05in,width=2.2in,angle=0]{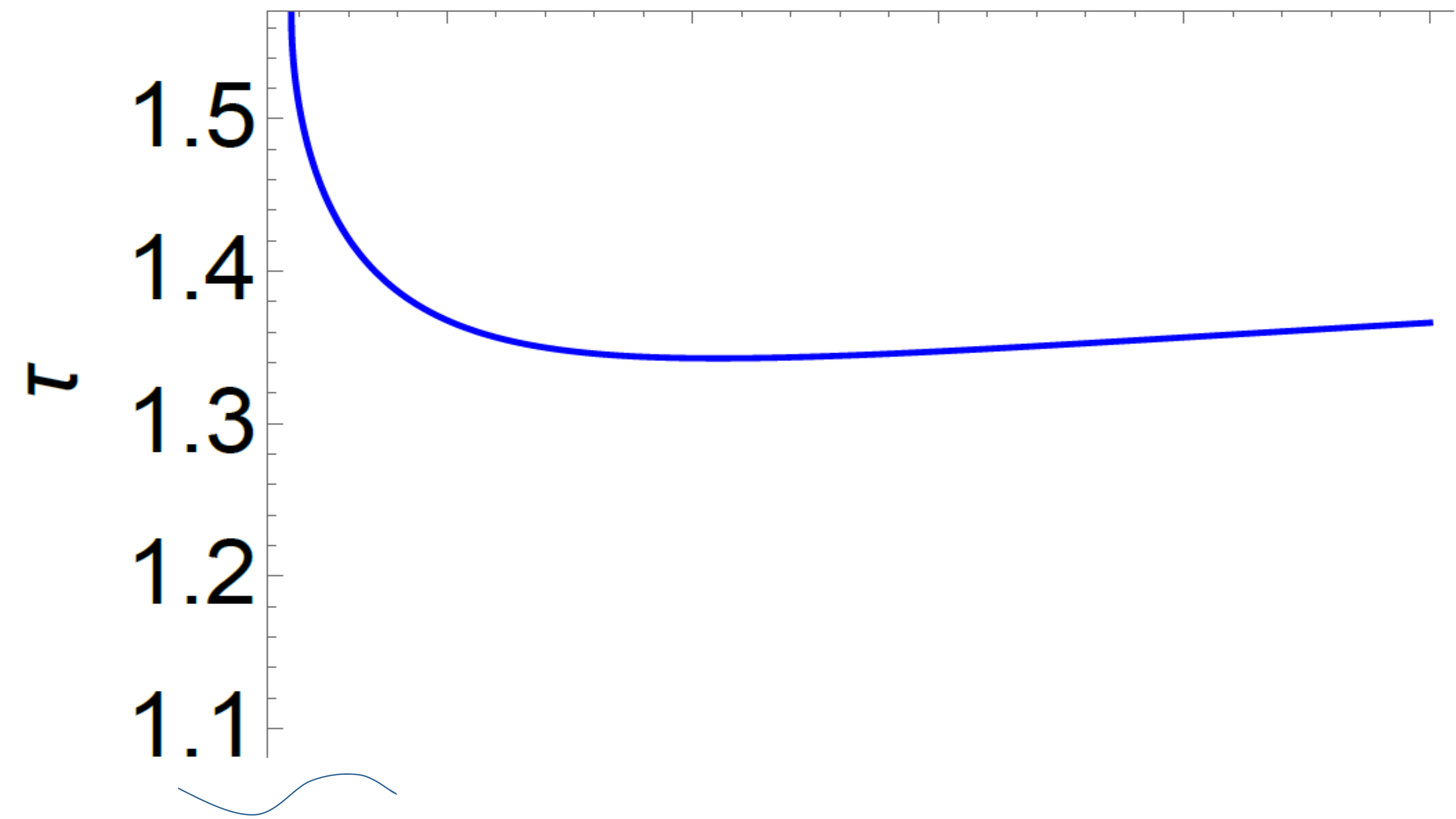}

  \includegraphics[height=1.24in,width=2.55in,angle=0]{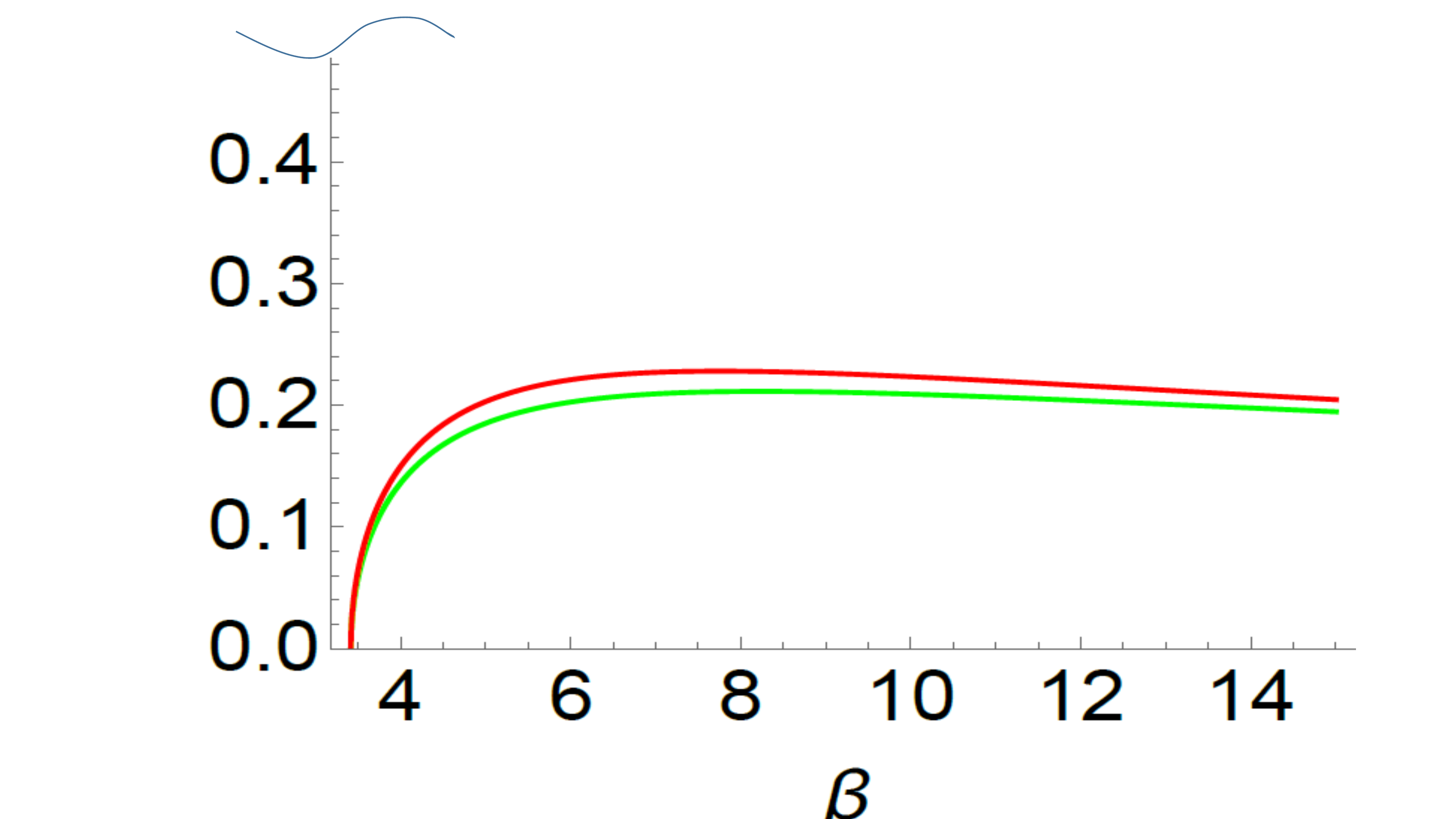}
 \caption{Signature changes in the induced metric  $g_{\mu\nu}$ and effective metric 
 $\gamma_{\mu\nu}$ for   deformed $CP^{2}$
 are given in the plot of $\tau$ versus $\tilde\beta$  in the region $ ii) \;3.41\,\lesssim\,\tilde\beta$.   A sign change in  $g_{\theta\theta}$  or $\gamma_{\theta\theta}$  is indicated by the green curve.
Sign changes   in  $g_{\tau\tau}$ or $ \gamma_{\psi\psi}$ are indicated by the red and blue curves. }
\label{fig:test}

\end{figure}

\subsubsection{Effective metric}
We next use (\ref{dfefmtrix}) to compute the effective metric $\gamma_{\mu\nu}$ for deformed $CP^2$.
Starting with the canonical Poisson brackets (\ref{pbsfrzzstr}), we now obtain the following results for the nonvanishing components of the symplectic matrix $[\Theta^{\mu\nu}]$:
\be \Theta^{\tau\psi}=\frac 1{\sin\tau\cos\tau}\qquad \Theta^{\theta\psi}=\frac{2\cot\theta}{\sin^2\tau}\qquad \Theta^{\theta\phi}=-\frac{2\csc\theta}{\sin^2\tau}\ee
Computing  determinants, we get
 \beqa\det\Theta &=&\frac{4\csc^2\theta}{\cos^2\tau\sin^6\tau}\cr&&\cr 
 |\det\gamma||\det\Theta|&=& |\det g||\det\Theta|= 4|g_{\tau\tau}|\,(\cos^2\tau-\mu^2\sin^2\tau)^2\eeqa
As a result,   the  nonvanishing components of the effective metric tensor are
\beqa \frac{ \gamma_{\tau\tau}}{ \sqrt{|\det\gamma||\det\Theta|}}&=&-1\cr&&\cr 
\frac{ \gamma_{\theta\theta}}{ \sqrt{|\det\gamma||\det\Theta|}}&=&\frac 1{4(\mu^2-\cot^2\tau)}
\cr&&\cr \frac{\gamma_{\psi\psi}}{ \sqrt{|\det\gamma||\det\Theta|}}&=&
\frac{1}{4(\mu ^2-\nu ^2 -{\rm sec}^2\tau\,{\rm csc}^2\tau) }\;,\label{fmt11}\eeqa
in addition to  $ \gamma_{\phi\phi}= \gamma_{\psi\psi}\,\cos^2\theta+ \gamma_{\theta\theta} \,\sin^2\theta$ and  $ \gamma_{\psi\phi}= \gamma_{\psi\psi}\,\cos\theta$.   As with the deformed $CP^{1,1}$ and $CP^{0,2}$ solutions, signature changes in the effective  metric coincide with signature changes in the induced  metric.
So like with the induced metric, the effective metric undergoes only one signature change  when  $\tilde\beta$ has the values in $i)$.    It is a change from the Euclidean signature to one where the effective metric has two space-like directions and two time-like directions. 

 Also like with the induced metric, the effective metric undergoes three signature changes when  $\tilde\beta$ has the values in $ii)$, which are  indicated in figure seven.
 A sign change in $\gamma_{\theta\theta}$ occurs for $\tan\tau=\frac 1{|\mu |}$ (indicated by the green curve in the figure), and sign changes in $\gamma_{\psi\psi}$ occur at  $\sin 2\tau=\frac 2{\sqrt{\mu^2-\nu^2}}$ (indicated by the red and blue curves in the figure).   The effective metric has  Euclidean signature below the red curve. In the tiny intermediate region  between the red and green curves, the effective metric, like the induced metric, has two-space-like directions and two time-like directions.  The effective metric  has Lorentzian signature in the intermediate region between the green and blue curves, with $\tau$ being time-like.  Above the blue curve, the induced metric has two-space-like directions and two time-like directions.

For the Lorentzian region, which we found    between the green and blue curves in figure seven, the effective metric describes a closed space-time cosmology.  For a fixed $\tilde\beta$ with values in $ii)$, the sign changes in $\gamma_{\psi\psi}$, depicted as red and  blue curves  in figure seven, correspond to space-time  singularities.  We denote the values of $\tau$ at these singularities by $\tau_0$ and $\tau_1$, with $\tau_0<\tau_1$.  Which one of these is the initial singularity, and which one is the final singularity, of course, depends on the direction of time. We obtain the time evolution of the spacial scale for this region in the next subsection.

\subsubsection{Expansion and contraction}

In the previous section, we saw that the effective metric for
 deformed  $CP^{2}$ can have  Lorentzian signature when   $\tilde\beta$ has the values in $ii)$.   In this case, $\tau$ is  the time-like coordinate, and it evolves from one signature change to another.   From (\ref{prtzseepee2}), $|\zeta_1|^2+|\zeta_2|^2=\tan^2\tau$, and so, like with the deformed $CP^{1,1}$ and $CP^{0,2}$ solutions,
a $\tau$-slice of the four-dimensional manifold away from singularities is a three-sphere, or more precisely, a Berger sphere.   We can compute 
the spatial scale $a(|\tau|)$ at any $\tau-$slice and proper time $t$ in the co-moving frame for the deformed  $CP^{2}$ solution, using (\ref{sptlscl}) and (\ref{comvngtm}), respectively. 
For the former, we get
\be a(|\tau|)^3=
{(4\pi)^2\,|\sin^4\tau\cos\tau|\,|\cot^2\tau-\mu^2|^{\frac 12}\;|(\mu^2-\nu^2)\sin^2\tau\cos^2\tau-1|^{\frac 14}}\ee  
It follows that the spatial scale vanishes at the signature changes, which are associated with the cosmological singularities.
For the  latter, (\ref{comvngtm}) gives
\beqa t(\tau)&=&2\,\int_{\tau_0}^{\tau}d\tau' \,|\cos^2\tau'-\mu^2\sin^2\tau'|^{\frac 12}\;\Big|(\mu^2-\nu^2)\sin^2\tau'\cos^2\tau'-1\Big|^{\frac 14}\eeqa
The lower integration limit $\tau_0$ corresponds to the value of $\tau$ at the coordinate singularity defined by  $\sin 2\tau_0=\frac 2{\sqrt{\mu^2-\nu^2}}$,  $\tau_0<\frac\pi  4$,
corresponding to the sign change in $\gamma_{\psi\psi}$.    In figure  eight, we plot $a(\tau)$ versus $t(\tau)$ for three values of $\tilde \beta$  in region $ii)$.   For  $\tau$ close to $\tau_0$, we get $a\sim (\tau-\tau_0)^\frac 1{12}$, $t\sim (\tau-\tau_0)^\frac 54$, and hence, $a\sim t^\frac1 {15}$. We find identical behavior near the other singularity  at  $\tau=\tau_1$.  We thus get a very rapid initial expansion and a very rapid final contraction.
\begin{figure}
  \centering
  \includegraphics[height=1.8in,width=2.in,angle=0]{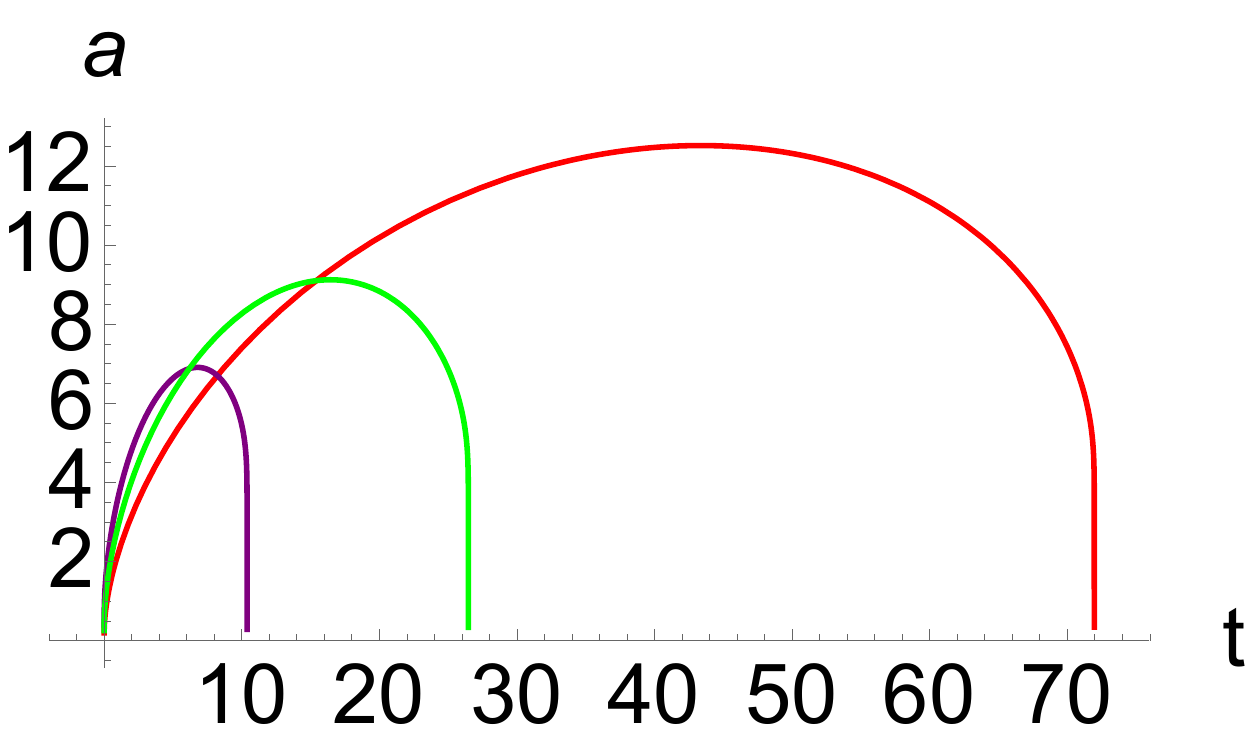}
  \caption{ $a(\tau)$ versus $t(\tau)$ for the region of deformed $CP^{2}$ where the effective metric has Lorentzian signature, for  $\tilde \beta=3.5$ (red curve), $3.75$ (green curve) and $5$ (purple curve). }
\label{fig:test}
\end{figure}

\section{Conclusions}
\setcounter{equation}{0}

We have obtained a number of new solutions to IKKT-type matrix models,  which exhibit signature change in the commutative limit.  All such examples found so far, including the non-commutative $H^4$ solution of \cite{Steinacker:2017vqw}, require including a mass term in the matrix action.  Since mass terms  result from an IR regularization,\cite{Kim:2011cr} it is interesting to speculate whether  signature change on the brane is connected to the regularization. On the other hand, we remark that the  mass term resulting from the regularization does not necessarily lead to signature change, since we have obtained solutions to the massive matrix model which exhibit no signature change in the commutative limit.  For example,  no sign changes occur in either the induced metric or effective metric for   deformed $CP^{0,2}$ when  $-1<\tilde\beta\lesssim -0.746$ and 
 $ -0.603\,\lesssim\,\tilde\beta\,\le -.6\,$.   Moreover, our work does not rule out the possibility of solutions to a {\it  massless} matrix model which exhibit  signature change in the commutative limit.

 The four-dimensional solutions of section five are deformations of non-commutative complex  projective spaces, specifically non-commutative $CP^2$, $CP^{1,1}$ and $CP^{0,2}$.  The manifolds that emerge from these solutions can have multiple signature changes. The manifolds resulting from  deformed  non-commutative   $CP^{0,2}$ solution can  undergo two signature changes, while those resulting from deformed non-commutative $CP^2$ solution can have up to three signature changes.
The regions where the effective metric of these manifolds have  Lorentzian signature serve as crude models of closed (in the case of non-commutative  $CP^2$) and open (in the case  of non-commutative $ CP^{1,1}$ and $CP^{0,2}$) cosmological space-times.  They contain cosmological singularities that are resolved away from the commutative limit.    The evolution of the spatial scale $a$  as a function of the proper time  $t$ in the co-moving frame  was computed for these examples.  For all examples (and also the example of non-commutative $H^4$ in   \cite{Steinacker:2017vqw}) an extremely rapid expansion (or contraction, in case of the big crunch singularity of the closed cosmology)  was found for the spacial scale $a$  near the cosmological singularities. Rather than following an  exponential  behavior, we obtained $a\sim t^\frac1 {15}$ near $t=0$ for  non-commutative $CP^2$ and $CP^{1,1}$, and  $a\sim t^\frac 19$ for  non-commutative $CP^{0,2}$. Also like  non-commutative $H^4$,\cite{Steinacker:2017vqw}  the space-times emerging from the deformed  non-commutative  $CP^{1,1}$ and $CP^{0,2}$  solutions expand linearly  at late times, $a\sim t$.  

 Unlike the space-time manifold that emerges from non-commutative $H^4$,\cite{Steinacker:2017vqw} the manifolds that emerge from non-commutative $CP^2$, $CP^{1,1}$ and $CP^{0,2}$ are not maximally symmetric.  For the latter manifolds, any time slice of the space-time is a Berger sphere.
Although being, perhaps, less realistic than  non-commutative $H^4$  with regards to cosmology, the  examples of non-commutative $CP^2$, $CP^{1,1}$ and $CP^{0,2}$ are considerably simpler spaces  than  non-commutative $H^4$,   with evidently similar outcomes for the evolution of the  spatial scale.   Non-commutative $H^4$ carries an additional bundle structure, that is not present for the solutions of section five.
In order to close the algebra  on non-commutative $H^4$, one must  extend it to a larger non-commutative space.  That space is non-commutative $CP^{1,2}$.  In the commutative limit, one recovers the $CP^{1,2}$ manifold, an $S^2$ bundle over $AdS^4$.

The eight-dimensional matrix model considered in sections four and five utilized a particular  indefinite background metric $\eta$,  the $su(2,1)$ Cartan-Killing metric.  Other indefinite  background metrics can be considered. $\eta$=diag($+,+,+,+,+,+,+,-)$ was used in \cite{Chaney:2016npa}, to obtain non-commutative $CP^2$ solutions.  We can preserve  $SO(3)$ rotational symmetry with a generalization of the background metric to $\eta$=diag($\kappa_3,\kappa_3,\kappa_3,-,-,-,-,\kappa_8)$,  $\kappa_3,\kappa_8=\pm$.  (We exclude $\kappa_3=\kappa_8=-$, since this will only produce a Euclidean induced and effective metric.) If for example, we search for deformed $CP^{1,1}$ and $CP^{0,2}$ solutions to the matrix equations (\ref{smclmtrxeq}) with this metric, then the conditions (\ref{cndtnsnmnab}) generalize to
\beqa ( 2\mu-\alpha)\Bigl(\mu^2\kappa_3+\frac 12\Bigr)+3\mu\tilde\beta&=&0\cr&&\cr
 \mu^2\kappa_3+\nu^2\kappa_8+2-\alpha(\mu\kappa_3+\nu\kappa_8)+4\tilde\beta&=&0\cr&&\cr
 2\nu-\alpha+2\nu \tilde\beta&=&0
 \eeqa
where we again assumed the ansatz (\ref{dfrmdnstz}).  Solutions for different choices of $\kappa_3$  and $\kappa_8$ may be found, although they may be quite nontrivial, and many more four-dimensional signature changing manifolds are expected to emerge in the commutative limit.

In this article we have neglected stability issues, and the addition of fermions.  The question of stable solutions to matrix models is highly non-trivial.  For two-dimensional solutions it was found previously that longitudinal and transverse fluctuations contribute with opposite signs to the kinetic energy.  It is unclear how the extension to a fully supersymmetric theory can resolve this issue.  We hope to address such questions in the future.
 
\bigskip
\appendice{\Large{\bf $\quad$Some properties of $su(2,1)$ in the defining representation}}

In terms of  $su(3)$ Gell-Mann matrices  $\lambda_a$,  the $su(2,1)$ Gell-Mann matrices  $\tilde \lambda_a$  are given by
\beqa \tilde\lambda_{\tt a}&=&\lambda_{\tt a} \;,\qquad\;\; {\tt a}=1,2,3,8 \cr &&\cr
 \tilde\lambda_{\tt a'}&=&i\lambda_{\tt a'}\;,\qquad {\tt a'}=4,5,6,7\eeqa
They satisfy the hermiticity properties (\ref{zroptone4}).

The structure constants for  $su(2,1)$ are   $C_{ab}^{\;\;\;c}=\tilde f_{abd}\eta^{dc}$, where $\eta_{ab}$ is the  Cartan-Killing metric (\ref{atedeemtrc}), and
  $\tilde f_{abc}$  are  totally antisymmetric, with the nonvanishing values
\beqa &&\tilde f_{123}=1\qquad\qquad\qquad\tilde f_{845}=\tilde f_{867}=-\frac {\sqrt{3}}2 \cr&&\cr
&& \tilde f_{147}= \tilde f_{165}=\tilde f_{246}=\tilde f_{257}=
\tilde f_{345}=\tilde f_{376}=-\frac 12\eeqa
Except for $\tilde f_{123}$ these structure constants are opposite in sign from those obtained from the standard Gell-Mann matrices  of $su(3)$.

 Some useful identities for  the $su(2,1)$ Gell-Mann matrices and $\tilde f_{abc}$  are
\beqa   {\rm tr}\,\tilde\lambda_a\tilde\lambda_b&=&[\tilde\lambda_a]^i_{\;\,j}[\tilde\lambda_b]^j_{\;\,i}\;=\;2\eta_{ab}\;,\\ &&\cr
  [\tilde\lambda_a,\tilde\lambda_b]_+&=&2\tilde d_{abc}\tilde\lambda^c+ \frac 43\eta_{ab} \BI  \\
&&\cr \tilde f_{abc}\tilde f^{bc}_{\;\;\;d}&=&3\eta_{ad}\label{tfsqeqeta}\\
&&\cr [\tilde\lambda^a]^{i}_{\;\,j} [\tilde\lambda_a]^{k}_{\;\,\ell}   &=& 2\delta^i_{\ell}\delta^k_{j}-\frac 23\delta^i_{j}\delta^k_{\ell}\label{frzidntee}\eeqa
$[\;,\;]_+$ denotes the anti-commutator, and   $\tilde d_{abc}$  are  totally symmetric, with the nonvanishing values
\beqa &&\tilde d_{443}=\tilde d_{553}=\tilde d_{146}=\tilde d_{157}=\tilde d_{256}=-\frac 12\qquad \tilde d_{663}=\tilde d_{773}=\tilde d_{247}=\frac 12\cr&&\cr &&\tilde d_{118}=\tilde d_{228}=\tilde d_{338}=\frac 1{\sqrt{3}}\qquad \tilde d_{448}=\tilde d_{558}=
\tilde d_{668}=\tilde d_{778}=\frac 1{2\sqrt{3}}\qquad\tilde d_{888}=-\frac 1{\sqrt{3}}\eeqa
(\ref{frzidntee}) is the Fierz identity, which has the same form as that for $su(3)$.

\bigskip
\appendice{\Large{\bf $\quad$Effective metric}}

Here we review the derivation of  (\ref{dfefmtrix}), relating  the effective metric $\gamma_{\mu\nu}$ to the induced metric $g_{\mu\nu}$.  We use the example of the massless scalar field.\cite{Steinacker:2010rh} 
We then apply the result to compute the effective metrics for (undeformed)
   $CP^{1,1}$ and  $CP^{0,2}$.

 Denote the scalar field by $\Phi=\Phi(X)$ on a non-commutative background spanned by matrices $X_a$.  The standard action is
\be -\frac 1{2k^2}{\rm Tr}[X_a ,\Phi][X^a \Phi]\;,\label{mslsslrfa}\ee
$a=1,2,...,d$.  Now take the semi-classical limit  $\hbar\rightarrow 0$.  This means again replacing matrices $X_a$ by commuting variables $x_a$, corresponding to embedding coordinates of  some continuous manifold.  $\Phi$ is then replaced by a function $\phi$ on the manifold, and commutators are replaced by $i\hbar$ times Poisson brackets. We also need  to replace the trace by an integration $\int d\mu(x)$, where $d\mu(x)$  is an invariant integration measure. Say that  the manifold is parametrized by $\sigma=(\sigma^1,\sigma^2,...,\sigma^n)$, $n\le d$, with symplectic two-form $\Omega=\frac 12 [\Theta^{-1}]_{\mu\nu} \,d\sigma^\mu\wedge d\sigma^\nu$.  Then  one can set $d\mu(x)=\frac {d\sigma}{\sqrt{|\det\Theta|}}$. Taking   $k\rightarrow \hbar \kappa$, the semi-classical limit of (\ref{mslsslrfa}) is
\beqa && -\frac 1{2{\kappa}^2}\int\frac {d\sigma}{\sqrt{|\det\Theta|}}\,\{x_a ,\phi\}\{x^a, \phi\}\cr&&\cr
&=& -\frac 1{2{\kappa}^2}\int\frac {d\sigma}{\sqrt{|\det\Theta|}}\,\Theta^{\rho\mu}\partial_\rho x_a \partial_\mu\phi\;\Theta^{\sigma\nu}\partial_\sigma x^a \partial_\nu\phi\cr&&\cr
&=& -\frac 1{2{\kappa}^2}\int\frac {d\sigma}{\sqrt{|\det\Theta|}}\,\Theta^{\rho\mu}g_{\rho\sigma }\Theta^{\sigma\nu}\,  \partial_\mu\phi\partial_\nu\phi\;\label{sclmslsslrfa}\eeqa
On the other hand, the standard action of a scalar field $\phi$ on a background metric $\gamma_{\mu\nu}$ is
\be  -\frac 1{2{\kappa}^2}\int d\sigma {\sqrt{|\det\gamma| }}\gamma^{\mu\nu}\partial_\mu \phi\partial_\nu\phi\ee
Identifying these two actions gives (\ref{dfefmtrix}).  

As examples, we  compute the effective metrics for (undeformed)
   $CP^{1,1}$ and  $CP^{0,2}$, and show  that they are identical to the corresponding induced metrics.
 \begin{enumerate}
 \item Effective metric for $CP^{1,1}\,.\;$ 
  Using (\ref{smplctoneone}), the nonvanishing components   $\Theta^{\mu\nu}$  for $CP^{1,1}$ are
\be  \Theta^{\tau\psi}=\frac 1{\cosh\tau\sinh\tau}\qquad \quad \Theta^{\theta\psi}=\frac {2\cot\theta}{\cosh^2\tau}\qquad\quad \Theta^{\theta\phi}=-\frac {2\csc\theta}{\cosh^2\tau} \label{nvrssmplctc11}\ee 
Then
 \be\det\Theta =\frac{4\csc^2\theta}{\sinh^2\tau\cosh^6\tau}\quad \qquad\quad |\det\gamma|= 4\cosh^6\tau\sinh^2\tau\sin^2\theta\label{detsfrcp11}\ee  Computing $\Theta^Tg\Theta$ we find the following nonvanishing components
$$[\Theta^Tg\Theta]^{\tau\tau} =-1\qquad\quad  [\Theta^Tg\Theta]^{\theta\theta}=\frac {4}{\cosh^2\tau}\qquad\quad  [\Theta^Tg\Theta]^{\phi\phi}=\frac {4\csc^2\theta}{\cosh^2\tau}$$
\be [\Theta^Tg\Theta]^{\psi\psi}=\frac {4(\cot^2\theta-{\rm csch}^2\tau)}{\cosh^2\tau}\qquad\quad [\Theta^Tg\Theta]^{\phi\psi}=-\frac {4\cot\theta\csc\theta}{\cosh^2\tau}\ee
 Using (\ref{dfefmtrix}) and (\ref{detsfrcp11}), we then get $\gamma_{\mu\nu}=g_{\mu\nu}$.
 
 \item Effective metric for $CP^{0,2}\,.\;$  Using (\ref{smplct2zero}) the nonvanishing components $\Theta^{\mu\nu}$ for $CP^{0,2}$ are
\be  \Theta^{\tau\psi}=\frac 1{\cosh\tau\sinh\tau}\qquad \quad \Theta^{\theta\psi}=\frac {2\cot\theta}{\sinh^2\tau}\qquad\quad \Theta^{\theta\phi}=-\frac {2\csc\theta}{\sinh^2\tau}\label{nverssmpzero2} \ee 
 Here \be\det\Theta =\frac{4\csc^2\theta}{\sinh^6\tau\cosh^2\tau}\quad \qquad|\det\gamma|=4\cosh^2\tau\sinh^6\tau\sin^2\theta\quad\label{detsfrcp02}\ee   
 The  nonvanishing components of
 $\Theta^Tg\Theta$ are
$$[\Theta^Tg\Theta]^{\tau\tau} =-1\qquad\quad  [\Theta^Tg\Theta]^{\theta\theta}=\frac {-4}{\sinh^2\tau}\qquad\quad  [\Theta^Tg\Theta]^{\phi\phi}=\frac {-4\csc^2\theta}{\sinh^2\tau}$$
\be [\Theta^Tg\Theta]^{\psi\psi}=\frac {-4(\cot^2\theta+{\rm sech}^2\tau)}{\sinh^2\tau}\qquad\quad [\Theta^Tg\Theta]^{\phi\psi}=\frac {4\cot\theta\csc\theta}{\sinh^2\tau}\ee
 Using (\ref{dfefmtrix}) and (\ref{detsfrcp02}), we once again get $\gamma_{\mu\nu}=g_{\mu\nu}$.
\end{enumerate}

\medskip
{\bf Acknowledgements}

A.S. is grateful for useful discussions with J. Hoppe, H. Kawai, S.~Kurkcuoglu and H. Steinacker, and wishes to thank  the members of the Erwin Schr\"odinger International Institute for their hospitality and support  during the Workshop on  “Matrix Models for non-commutative Geometry and String Theory”.

\end{document}